\documentclass[%
 reprint,
 amsmath,amssymb,
 aps,
]{revtex4-2}

\usepackage{graphicx}
\usepackage{dcolumn}
\usepackage{bm}
\usepackage{color}
\usepackage{float}

\begin{document}

\preprint{APS/123-QED} 

\title{ Anisotropic fluid spheres in the framework of $f(R,\mathcal{T})$ gravity theory}

\author{S. K. Maurya} \email{sunil@unizwa.edu.om}
 \affiliation{ Department of Mathematics and Physical Science,
College of Arts and Science, University of Nizwa, Nizwa, Sultanate of Oman}

\author{Francisco Tello-Ortiz} \email{francisco.tello@ua.cl}
\affiliation{Departamento de F\'isica, 
Facultad de ciencias b\'asicas, Universidad de Antofagasta, Casilla 170, 
Antofagasta, Chile}

\date{\today}

\begin{abstract}
The main aim of this paper is to obtain analytic relativistic anisotropic spherical solutions in f(R,$\mathcal{T}$) scenario. To do so we use modified Durgapal-Fuloria metric potential and the isotropic condition is imposed in order to obtain the effective anisotropic factor $\tilde{\Delta}$. Besides, a notable and viable election on f(R,$\mathcal{T}$) gravity formulation is taken. Specifically $f(R,\mathcal{T})=R+2\chi\mathcal{T}$, where $R$ is the Ricci scalar, $\mathcal{T}$ the trace of the energy-momentum tensor and $\chi$ a dimensionless parameter. This choice of $f(R,\mathcal{T})$ function modifies the matter sector only, including new ingredients to the physical parameters that characterize the model such as density, radial, and tangential pressure. Moreover, other important quantities are affected such as subliminal speeds of the pressure waves in both radial and transverse direction, observational parameters, for example, the surface redshift which is related with the total mass $M$ and the radius $r_{s}$ of the compact object. Also, a transcendent mechanism like equilibrium through generalized Tolman-Oppenheimer-Volkoff equation and stability of the system are upset. We analyze all the physical and mathematical general requirements of the configuration taking $M=1.04 M_{\odot}$ and varying $\chi$ from $-0.1$ to $0.1$. It is shown by the graphical procedure that $\chi<0$ yields to a more compact object in comparison when $\chi\geq0$ (where $\chi=0.0$ corresponds to general relativity theory) and increases the value of the surface redshift. However, negative values of $\chi$ introduce in the system an attractive anisotropic force (inward) and the configuration is completely unstable (corroborated employing Abreu's criterion). Furthermore, the model in Einstein gravity theory presents cracking while for $\chi>0$ the system is fully stable. The relationship between effective pressures and effective density  $\tilde{\rho}$ is discussed and obtained. This is achieved by establishing the corresponding equation of state.
\end{abstract}

\maketitle

\section{Introduction}
Put forward by Harko and his collaborators \cite{r1}, f(R,$\mathcal{T}$) gravity theory was designed to face the late-time acceleration of the Universe and the existence of dark matter. All these issues provided by recent observational data \cite{r2,r3,r4,r5,r6}. At present f(R,$\mathcal{T}$) theory is an active research field in the cosmological context \cite{r8,r9,r10,r11,r12,r13,r14}. Other interesting works available in the literature are for example the study by Sharif et.al about the non-static line element for collapsing of a spherical body having anisotropic fluid \cite{r15}, the static spherical wormhole solutions found in \cite{r16,r17}. Moreover, perturbation techniques were used by Bhatti et.al in the study of spherical stars \cite{r18}. The effects on gravitational lensing due to f(R,$\mathcal{T}$) gravity were discussed by Houndjo in \cite{r19}. Furthermore, Baffou et al. \cite{r20} employed perturbation on de-Sitter space-time and power-law models in order to explore some cosmic viability bounds.\\
Even though the study of the Universe as a whole is a very intriguing current problem of dealing with, studies and investigations of structures within it such as neutron stars, white dwarfs and black holes among others, which constitute real laboratories to analyze in a fragmented way the most hidden secrets of the Universe, it is expected once these secrets are revealed, they provide us with the expected response that will finally put the pieces of this great puzzle in the corresponding place. In this direction, many authors have investigated the existence of collapsed structures within the framework of  f(R,$\mathcal{T}$) theory,
exploring how different models of f(R, $\mathcal{T}$) affect on the principal properties of these kind of objects, besides contrasting the reported results with general relativity theory (GR hereinafter)\cite{r21,r22,r23,r24,r25,r26,r27,r28}. Of course, is not an easy task solve f(R, $\mathcal{T}$) field equations (as in GR case), for this reason one needs to prescribe additional information such as a suitable metric potential, an electric field (in the corresponding case), an adequate anisotropy factor or an equation of state (EoS from now on). Concerning the latter, the obtaining and not imposing of the EoS leads to a better understanding of how the matter confined in the stellar interior behaves. Moreover, it is possible to determine what type of material constitutes the distribution, for example, ordinary matter such as neutrons or strange matter such as quarks \cite{harko1} (and references contained therein). Additionally, the EoS provides the relation between the macro observables parameters of the star such as mass and radius.
Following the same spirit, in this work we study the existence of compact objects, specifically neutron stars (or the possibility of quark stars also known as strange stars) in the framework of f(R,$\mathcal{T}$) gravity. The main goal is to build up the full geometrical description of the interior space-time taking as departure point Durgapal-Fuloria \cite{Durgapal} metric potential $\xi$ and introduce an anisotropic behavior of the matter distribution. The latter is achieved imposing the isotropic condition at the level of the field equations considering $p_{r}\neq p_{t}$ once a viable form of the f(R,$\mathcal{T}$) function is chosen. In this opportunity we have selected the modified gravity model to be $f(R,\mathcal{T})=R+2\chi\mathcal{T}$ \cite{r1}, being $R$ the Ricci scalar, $\mathcal{T}=g^{\mu\nu}T_{\mu\nu}$ the trace of the energy-momentum tensor (As was pointed out by Harko et.al \cite{r1} the dependence from $\mathcal{T}$ may be induced by exotic imperfect fluids or quantum effects) and $\chi$ a coupling dimensionless constant. This coupling constant in some sense quantifies the effect on the matter sector modifications and in the geometrical one also (it is included in the Durgapal-Fuloria ansatz).   \\
The study of compact objects driven by anisotropic matter distributions must fulfill some general and basic requirements in order to be a physical and mathematical admissible model from the astrophysical point of view. Taking into account such conditions, with the support of graphic analysis we have studied and corroborated the fulfillment of each of them, which have been established throughout history from the pioneering work by Bowers and Liang \cite{r29}, in research developed by Herrera, Ponce de Le\'on, Cosenza, Di Prisco and Ivanov, to name a few \cite{r30,r31,r32,r33,r34,r35,r36,r37,r38,r39,r40,r41,r42,r43,r44,r45}. Although these demands have been widely developed and applied to the study of collapsed structures within the framework of GR \cite{r46} (and references contained therein). The same in general must be satisfied in the scenario of modified gravity theories, since these theories under some limit must reproduce GR and its results, which are known to be very precise and have been verified several times. Of course, the chosen f(R,$\mathcal{T}$) model in this research reproduces the original Durpgapal-Fuloria solution developed in the context of GR in the limit $\chi=0.0$ and $\alpha=1$.\\ 
So, the obtained model has interesting properties that can be compared with the obtained ones in the GR frame. In considering $-0.1\leq\chi\leq0.1$ and $M=1.04M_{\odot}$ (regarding also the GR limit \i.e $\chi=0.0$) we have performed a full study and checked all the necessary and sufficient conditions in order to describe an acceptable compact object. So, in section \ref{2} we derive the complete set of equations corresponding to f(R,$\mathcal{T}$) theory, and the matter content is fixed \i.e the energy-momentum tensor $T_{\mu\nu}$ and the Lagrangian matter $\mathcal{L}_{m}$. In Sec. \ref{3} the full model has specified \i.e the inner space-time geometry and the material content. In Sec. \ref{4} the internal geometry is smoothly joining with the exterior Schwarzschild geometry. In Secs. \ref{5} and \ref{6} we analyze the full main salient features of the obtained model. Studying the physical and mathematical behavior of the space-time geometry and all the thermodynamic variables such as the effective radial and tangential pressure, the effective energy density. In Sec. \ref{ener} by means of energy conditions we check if the energy-momentum tensor is driven the material content if well behaved at every point within the star. In Sec. \ref{7} causality condition and its implications are discussed.
Furthermore, we discuss the influence of anisotropies fluids distributions and f(R,$\mathcal{T}$) model on the surface redshift. Sec. \ref{5.2} is devoted to the analysis of the Tolman-Oppenheimer-Volkoff equilibrium equation in the f(R,$\mathcal{T}$) context. In order to explore how the whole system is affected by the different forces. We provide in Sec. \ref{9} the corresponding equation of state, examining its properties and discussing its importance in the study of collapsed configurations such as neutron stars. In Sec. \ref{10} the stability of the system is analyzed using Abreu's criteria. Finally, in Sec. \ref{11} we reinforce all the good properties obtained in this work and comparisons between f(R,$\mathcal{T}$) gravity theory and GR are given.

\section{A Formulation for Modified $f(R, \mathcal{T}$) Gravity Theory }\label{2}
 Let us consider the integral action $S$ for modified $f(R,\mathcal{T})$ gravity theory as
\begin{eqnarray}\label{1.1}
S=\frac{1}{16\pi}\int f(R,\mathcal{T})\sqrt{-g} d^{4}x+\int \mathcal{L}_{m}\sqrt{-g}d^{4}x, 
\end{eqnarray} 
where relativistic geometrized units were employed i.e $c=G=1$. Here, $f(R,\mathcal{T})$ is an arbitrary function of the Ricci scalar $R$
and the trace $\mathcal{T}$ of the energy-momentum tensor $T_{\mu\,\nu}$, while $\mathcal{L}_{m}$ denotes the matter Lagrangian density.   \\  
By variation of action $S$ with respect to the metric tensor $g_{\mu\nu}$ yields the following field equation 
\begin{eqnarray}\label{1.2}
\left( R_{\mu\nu}- \nabla_{\mu} \nabla_{\nu} \right)f_R (R,\mathcal{T}) +\Box f_R (R,\mathcal{T})g_{\mu\nu} - \frac{1}{2} f(R,\mathcal{T})g_{\mu\nu}   \nonumber \\ = 8\pi T_{\mu\nu}  - f_\mathcal{T}(R,\mathcal{T})\, \left(T_{\mu\nu}  +\Theta_{\mu\nu}\right)~~~~~
\end{eqnarray} 

where $f_R (R,\mathcal{T})$ denote the partial derivative of $f (R,\mathcal{T})$ with respect to $R$ and $f_\mathcal{T} (R,\mathcal{T})$ is the partial derivative of $f (R,\mathcal{T})$ with respect to $\mathcal{T}$, while $R_{\mu\nu}$ is the Ricci tensor. The box operator ${\Box \equiv \partial_{\mu}(\sqrt{-g} g^{\mu\nu} \partial_{\nu})/\sqrt{-g}}$ is called the D'Alambert operator, and $\nabla_\mu$ represents the
covariant derivative associated with the Levi-Civita connection of metric tensor $g_{\mu\nu}$. 
The stress energy tensor $T_{\mu\nu}$ and $\Theta_{\mu\nu}$ are defined as follows, 
\begin{eqnarray}
T_{\mu\nu}&=&g_{\mu\nu}\mathcal{L}_m-2\partial\mathcal{L}_m/\partial g^{\mu\nu},\\
\Theta_{\mu\nu}&=& g^{\alpha\beta}\delta T_{\alpha\beta}\,/\,\delta g^{\mu\nu}. 
\end{eqnarray}

Using Eq.(\ref{1.2}), the Einstein tensor $G_{\mu\nu}$ can be written as,
\begin{eqnarray}\label{eins}
\resizebox{0.98\hsize}{!}{$G_{\mu\nu}=\frac{1}{f_{R}\left(R,\mathcal{T}\right)}\bigg[8\pi\,T_{\mu\nu}+\frac{1}{2}\left(f\left(R,\mathcal{T}\right)-Rf_{R}\left(R,\mathcal{T}\right)\right)g_{\mu\nu}$}  \nonumber \\ \resizebox{.93\hsize}{!}{$-T_{\mu\nu}\,f_{\mathcal{T}}\left(R,\mathcal{T}\right)-\left(g_{\mu\nu}\Box-\nabla_{\nu}\nabla_{\mu}\right)f_{R}\left(R,\mathcal{T}\right) \bigg].$} ~~~  
\end{eqnarray} 
After taking the covariant derivative of the Eq.(\ref{1.2}), we get

\begin{eqnarray}\label{1.3}
\nabla^{\mu}T_{\mu\nu}=\frac{f_\mathcal{T}(R,\mathcal{T})}{8\pi -f_\mathcal{T}(R,\mathcal{T})}\bigg[(T_{\mu\nu}+\Theta_{\mu\nu})\nabla^{\mu}\ln f_\mathcal{T}(R,\mathcal{T})  \nonumber \\+\nabla^{\mu}\Theta_{\mu\nu}-\frac{1}{2}g_{\mu\nu}\nabla^{\mu}\mathcal{T} \bigg]~~~~~
\end{eqnarray}

The equation (\ref{1.3}) shows that the covariant derivative of stress-energy momentum tensor $T_{\mu\nu}$ not vanishes in modified $f(R,\mathcal{T})$ gravity as in other theories of gravity. Throughout in our study,
we consider the Lagrangian matter  $\mathcal{L}_m=-\mathcal{P}$ where $\mathcal{P}=\frac{1}{3}(p_r+2p_t)$\cite{DebMNRS}. Then using Eq.(4) we obtain $\Theta_{\mu\nu}=-2T_{\mu\nu}-\mathcal{P} g_{\mu\nu}$.

In order to obtain the effective stress-energy momentum tensor for modified theory of gravity  we consider the simplest linear functional form of $f(R,\mathcal{T})$ (proposed by Harko et al. \cite{r1}) as follows 
\begin{eqnarray}
f(R,\mathcal{T})=R+2\chi\mathcal{T}
\label{frt}
\end{eqnarray}  
where $\chi$ is a coupling constant. The above $f(R,\mathcal{T})$ function has been used widely used to develop the different $f(R,\mathcal{T})$ gravity compact objects. By inserting the value $f(R,\mathcal{T})$ from Eq. (\ref{frt}) in Eq. (\ref{eins}) we obtain
\begin{eqnarray}\label{1.5}
G_{\mu\nu}=8\pi T_{\mu\nu}+\chi \mathcal{T}g_{\mu\nu}+2\chi(T_{\mu\nu}+\mathcal{P} g_{\mu\nu})\nonumber\\=8\pi \tilde{T}_{\mu\nu},~~~   
\end{eqnarray}

Here, we consider the energy momentum tensor $T_{\mu\nu}$ corresponding to anisotropic fluid distribution which can be defined as,
\begin{equation}\label{1.4}
T_{\mu\nu}=(\rho+{p_t})u_\mu u_\nu-{p_t}g_{\mu\nu}+\left({p_r}-{p_t}\right)v_\mu v_\nu,
\end{equation}
where ${u_{\nu}}$ is the four velocity, satisfying $u_{\mu}u^{\mu}=-1$ and $u_{\nu}\nabla^{\mu}u_{\mu}=0$. Here, $\rho$ is matter density,  while $p_r$ and $p_t$ are radial pressure and tangential pressure respectively. 
Now the modified energy-momentum tensor $\tilde{T}_{\mu\nu}$ can be written as  
\begin{eqnarray}\label{1.5a}  
\tilde{T}_{\mu\nu}= {T}_{\mu\nu}\,\left(1+\frac{\chi}{4\pi}\right)+\frac{\chi}{8\pi}(\mathcal{T}+2\mathcal{P})g_{\mu\nu}\label{1.5aa}~~
\end{eqnarray} 
 
By inserting the value of $f(R,\mathcal{T})=R+2\chi\mathcal{T}$ in Eq.(\ref{1.3}) we get 
\begin{eqnarray}\label{1.6a}
\nabla^{\mu}T_{\mu\nu}=-\frac{1}{2\,\left(4\pi+\chi\right)}\chi\bigg[g_{\mu\nu}\nabla^{\mu}\mathcal{T}+2\,\nabla^{\mu}(\mathcal{P}\, g_{\mu\nu})].
\end{eqnarray}
By using the Eqs.(\ref{1.5a}) and (\ref{1.6a}) we can write,
\begin{eqnarray}
\nabla^{\mu}\tilde{T}_{\mu\nu}=0\label{1.6aa}
\end{eqnarray}

\section{The Field Equation for anisotropic matter distributions in modified $f(R,\mathcal{T})$ gravity}\label{3}
Let us consider the spacetime being static and spherically symmetric, which describes the interior of the object can be written in the following form
\begin{equation}\label{metric1}
ds^{2} = -e^{\nu(r) } \, dt^{2}+\xi^{-1} dr^{2} +r^{2}(d\theta ^{2} +\sin ^{2} \theta \, d\phi ^{2}),
\end{equation}

Since the modified energy tensor $\tilde{T}_{\mu\nu}$ is sum $\theta_{\mu\nu}$ which will clearly generates the anisotropic pressure within the effective matter distribution. Hence, using Using Eqs. (\ref{1.5}) and (\ref{1.5aa}) together with line element (\ref{metric1}) The field equations for the spherically symmetric anisotropic stellar system can be written as,

\begin{eqnarray}\label{effectivedensity}
8\pi\,\tilde{\rho}  &=&\bigg[\frac{1}{r^{2}}-\frac{\xi}{r^{2}}-\frac{\xi^{\prime}}{r}\bigg]\label{2.1a}~~~\\ 
8\pi\, \tilde{p}_r &=&\bigg[\xi\left(\frac{1}{r^{2}}+\frac{\nu^{\prime}}{r}\right)-\frac{1}{r^{2}}\bigg]\label{p2}\label{2.1b}\\
8\pi\,\tilde{p}_t&=&\bigg[\frac{\xi}{4}\left(2\nu^{\prime\prime}+\nu^{\prime2}+2\frac{\nu^{\prime}}{r}\right)+\frac{\xi^{\prime}}{4}\left(\nu^{\prime}+\frac{2}{r}\right)\bigg].\label{2.1c}~~~~~
\end{eqnarray} 
The primes denote differentiation with respect to the radial coordinate $r$. Using the Eqs.(\ref{1.5a}) and (\ref{1.5aa}) the effective quantities like effective density ($\tilde{\rho}$), effective radial pressure ($\tilde{p}_r$) and effective tangential pressure ($\tilde{p}_r$) can be written in terms of anisotropic pressures $p_r$ and $p_t$, energy density $\rho$ for anisotropic matter distribution as,  
\begin{eqnarray}
8\,\pi\,\rho+{\frac {\chi\,}{3}}\left( 9\,\rho-p_{{r}}-2\,p_{{t}} \right)&=&8\,\pi\,\tilde{\rho}\label{2.2a}\\
8\,\pi\,p_{{r}}-{\frac {\chi\,}{3}}\left( 3\,\rho-7\,p_{{r}}-2\,p_{{t}} \right)&=&8\,\pi\,\tilde{p}_r,\label{2.2b}\\
8\,\pi\,p_{{t}}-{\frac {\chi\,}{3 }}\left( 3\,\rho-p_{{r}}-8\,p_{{t}} \right) &=&8\,\pi\,\tilde{p}_t,\label{2.2c}
\end{eqnarray}
Eqs.(\ref{2.1a}) defines the gravitational mass inside the star of radius $r$. Using the Eqs.(\ref{2.1a}) and (\ref{2.2a}) with $\xi=1-\frac{2 m(r)}{r}$ we get,
\begin{eqnarray}\label{eq20}
\tilde{m} =4\pi\int_0^r{ r^2 \tilde{\rho}\,dr} 
\end{eqnarray}
For this modified anisotropic matter distribution, it is also necessary that the anisotropic fluid must satisfy another additional equation (known as conservation equation) as,
\begin{eqnarray}
p^{\prime}_r+\frac{\nu^{\prime}}{2}\,(\rho+p_r)-\frac{2}{r}\,(p_t-p_r)=\frac{\chi\,(3\rho^{\prime}-p^{\prime}_r-2p^{\prime}_t)}{6\,(4\,\pi+\chi)}.~~~~ \label{TOV} 
\end{eqnarray} 
The above conservation Eq.(\ref{TOV}) is also known as modified Tolman-Oppenheimer-Volkoff (TOV) equation for modified theory of gravity. It is note that this modified conservation equation reduce into conservation equation for General relativity when $\chi=0.0$.  On the other hand, the energy density ($\rho$), radial pressure ($p_r$) and tangential pressure ($p_t$) for anisotropic stellar model in modified $f(R,\mathcal{T})$ gravity theory can be written as, 
\begin{eqnarray}
\rho&=&\frac{8\,\pi\,\tilde{\rho}}{(8\pi+4\chi)}+\frac{8\,\pi\,(3\tilde{\rho}+\tilde{p_r}+2\tilde{p}_t)\,\chi}{3\,(8\pi+2\chi)\,(8\pi+4\chi)}\\
p_r&=&\frac{8\,\pi\,\tilde{p_r}}{(8\pi+2\chi)}+\frac{8\,\pi\,(3\tilde{\rho}-\tilde{p_r}-2\tilde{p}_t)\,\chi}{3\,(8\pi+2\chi)\,(8\pi+4\chi)}\\
p_t&=&\frac{8\,\pi\,\tilde{p_t}}{(8\pi+2\chi)}+\frac{8\,\pi\,(3\tilde{\rho}-\tilde{p_r}-2\tilde{p}_t)\,\chi}{3\,(8\pi+2\chi)\,(8\pi+4\chi)} 
\end{eqnarray}
\subsection{{Solution of field Equations (\ref{2.1a})- (\ref{2.1c}) in $f(R, \mathcal{T})$ gravity theory:}}  

To solve the equations (\ref{2.1a}), (\ref{2.1b}) and (\ref{2.1c}), we use the isotropy condition in Eqs. (\ref{2.1b}) and (\ref{2.1c}) which leads a second order differential equation of the form as, 
\begin{eqnarray}
{\xi}\left(\frac{\nu^{\prime\prime}}{2}+\frac{\nu^{\prime2}}{4}-\frac{\nu^{\prime}}{2\,r}-\frac{1}{r^2}\right)&+&\frac{\xi^{\prime}\nu^{\prime}}{4}+\frac{2\,\xi^{\prime}}{r}-\frac{1}{r^2}\\ = 8\pi(\tilde{p}_t-\tilde{p}_r)&=&(8\pi+2\,\chi)\,(p_t-p_r),~\label{diff} 
\end{eqnarray}
The above pressure isotropy equations in $f(R,\mathcal{T})$ gravity will be same as in GR if $\chi=0$. Since  Eq.(\ref{diff}) contain three unknowns $\nu$, $\xi$ and $\Delta=p_t-p_r$. Then to solve this equation, we choose a modified anstaz of gravitational potential $\xi$, proposed by Durgapal-Floria \cite{Durgapal}, of the form as,
\begin{eqnarray}
\xi&=&1-\frac{8\,x\,(3+x)}{7\,(1+\chi+x)^2},\label{garv1}
\end{eqnarray}

The choice of above metric potential (\ref{garv1}) is well motivated, because it is free from physical and mathematical singularities everywhere within the compact structure. Furthermore, yields to a finite, well defined and decreasing outward energy density at all points inside the star. The inclusion of $\chi$ in potential $\xi$ will cause the effect in modified energy density ($\tilde{\rho}$).  Now it can be noted that if $\chi=0$ and $p_t=p_r$, then $\nu=4\,\ln(1+A\,r^2)$ will satisfy the isotropy Eq.(\ref{diff}). By keeping this point in our mind, we construct the expression for $\Delta$, by using isotropy condition (\ref{diff}) and potential $\xi$, of the form as,   
\begin{eqnarray}
\Delta&=&\frac{8\,x\,\big[5 - \chi + x + 2 \alpha\,\Delta_1(x) + 
   \alpha^2 \big(\Delta_2(x) + \chi\, \Delta_3(x)\big)\big]}{7\, (1 + \chi + x)^3\,(1 + \alpha\, x)^2\,(8\pi+2\,\chi)}.~~~~~~\label{aniso} 
\end{eqnarray}
with~~ $\Delta_1(x)=[(-6 + 7\, x + x^2 - \chi\, (6 + 5\, x))]$,\\
$\Delta_2(x)=[\,7 + 7 \chi^3 - 15\, x - 2\, x^2 + 21\,\chi^2\, (1 + x)\,]$,\\
~~~$\Delta_3(x)=[\,(21 + 6\, x + 4\, x^2)\,]$. \\  

It is observe from Eq. (\ref{aniso}), the anisotropy $\Delta$ is zero at centre, and then $p_t=p_r$ at centre. However, the other details of physical features for $\Delta$ has been discussed in Sec. (V). Now by substituting the gravitational potential $\xi$ and anisotropic factor $\Delta$ from Eqs.(\ref{garv1}) and (\ref{aniso}) into Eq.(\ref{diff}), and by using the transformation $\nu=2\ln \Psi$ and $x=Ar^2$  we obtain,
\begin{widetext}
\begin{equation}
 \frac{d^2\Psi}{dx^2}+\frac{ 4  \big[3 - x + \chi\,(3 + 2\,x)\big]}{(1 + \chi + x)\,\big[7 + 7 \chi^2 - 10 x - x^2 + 14 \chi (1 + x)\big]}\,  \frac{d\Psi}{dx}-\frac{2\,(1 + \alpha\,x)^{-2}\, \big[-4\,\alpha\, (3 - x + 3\,\chi+ 2 \chi\,x) + \alpha^2\,f (x)\big] }{(1 + \chi + x)\,\big[7 + 7 \chi^2 - 10 x - x^2 + 14 \chi (1 + x)\big]}\,\Psi=0.  
 \label{Diff2}
\end{equation}

where,~~ $f(x)=\big[7 + 7 \chi^3 - 15 x - 7 x^2 - x^3 + 21 \chi^2 (1 + x) + \chi (21 + 6 x + 5 x^2)\big]$. 
It is note that the value $\Psi=(1+\alpha\,x)^2$ satisfy the above differential Eq.(\ref{Diff2}) which implies that this value of $\Psi$ leads a particular solution of Eq.(\ref{Diff2}). Then most general solution of Eq.(\ref{Diff2}) is given (using the change of dependent variable method) as,

\begin{eqnarray}
\Psi(x)=\Psi_{1}(x)\Bigg[C+D \Bigg(\frac{\sqrt{7}\,\sqrt{\Psi_ 2 (x)} \big[f_ 1 (x) + f_ 2 (x)\big]}{3\, \Psi_ 3 (x)\,\Psi_1(x)\,\sqrt{\Psi_1(x)}} +\frac{\sqrt{7}\,f_3 (x) }{(\Psi _ 3 (x))^{7/2}}\,\ln\bigg[\frac {-a (\Psi_3 (x))^{5/2}\, f_ 5 (x)} { 2 \, 
     f_ 4 (x)\, (1 + a\,x)}\bigg]\Bigg)\Bigg],~~~~~\label{Sol1}   
\end{eqnarray}      
where $C$ and $D$ are arbitrary constant of integration and expression of the used coefficients are as follows:  \\\\
$\Psi_ 1 (x) = (1 + \alpha\, x)^2,~~~~\Psi_ 2 (x) = \big[7 + 7 \chi^2 - 10 x - x^2 + 14 \chi (1 + x)\big],~~\Psi_ 3 (x) = \big[-1 + 7 \alpha^2 (1 + \chi)^2 - 2 \alpha (-5 + 7 \chi)\big],\\\\
f_ 1 (x) = (1 - \alpha - \alpha\,\chi) \Psi^2_ 3 (x) + \Psi_ 3 (x)\, \big[-1 + 6 \alpha \chi + \alpha^2 (-23 - 16 \chi + 7 \chi^2)\big] (1 + \alpha x),\\\\
f_ 2 (x) = \big[1 + \alpha (7 - 23 \chi) + \alpha^2 (63 + 174 \chi - 105 \chi^2) + \alpha^3 (-359 + 201 \chi + 399 \chi^2 - 161 \chi^3)\big]\,(1 + \alpha\,x)^2,\\\\
f_ 3 (x) = 4 \big[-1 + 2 \chi + \alpha (-1 - 23 \chi + 14 \chi^2) + \alpha^2 (37 + 36 \chi - 147 \chi^2 + 70 \chi^3) + \alpha^3 (-163 + 257 \chi + 21 \chi^2 - 301 \chi^3 + 98 \chi^4)\big],\\\\
f_ 4 (x) = \big[-1 + 2 \chi + \alpha (-1 - 23 \chi + 14 \chi^2) + \alpha^2 (37 + 36 \chi - 147 \chi^2 + 70 \chi^3) + 
    \alpha^3 (-163 + 257 \chi + 21 \chi^2 - 301 \chi^3 + 98 \chi^4)\big],\\\\
f_ 5 (x) = 5 - 7 \chi + x + \sqrt{\Psi_ 3 (x)}\, \sqrt{\Psi_ 2 (x)} + \alpha \big[7 + 7 \chi^2 - 5 x + 7 \chi (2 + x)\big]$. \\ 
\end{widetext}

By plugging the value of $\xi$, $\nu=2\ln \Psi$ from Eqs. (\ref{garv1}), and (\ref{Sol1}) into Eqs.(\ref{2.1a})- (\ref{2.1c}) we find the effective energy density $\tilde{\rho}$, effective radial pressure ($\tilde{p}_r$) and effective tangential pressure ($\tilde{p}_t$) in $f(R,\mathcal{T})$ gravity theory,

\begin{eqnarray} 
\tilde{\rho}&=&\frac{8\,\big[9 + 2\,x + x^2 + \chi (9 + 5\,x)\big]}{56 \pi \,(1 + \chi + x)^3}\label{50},\\
\tilde{p}_r&=&\frac{A\,\big[-8\,(3+x)+4\,\Psi_2(x)\,\Psi_4(x)\big]}{56 \pi \,(1+\chi+x)^2},\label{51}
\end{eqnarray} 
\begin{eqnarray}
\tilde{p}_{t}&=&\frac{A\,\big[-8\,(3+x)+4\,\Psi_2(x)\,\Psi_4(x)\big]}{56 \pi \,(1+\chi+x)^2}\nonumber\\&& +
\frac{8\,A\,x\,\big[ \Psi_5(x)+ \Psi_6(x)+\chi\,(21 + 6\, x + 4\, x^2)\big]}{56\,\pi\, (1 + \chi + x)^3\,(1 + \alpha x)^2}.~~~~~ \label{52}
\end{eqnarray} 
where, 
\begin{eqnarray}
\Psi_4(x)&=&\frac{\sqrt{7}\,D\, (1 +\chi+x)}{\Psi_1(x)\, \Psi(x)\,\sqrt{\Psi_2(x)}} + \frac{2\,\alpha}{\sqrt{\Psi_1(x)}}, \nonumber\\
\Psi_5(x)&=& \alpha^2\,[5 - \chi + x+2 \alpha\,(-6 + 7\, x + x^2 - \chi\, (6 + 5\, x))], \nonumber\\
\Psi_6(x)&=& \alpha^2\,[\,7 + 7 \chi^3 - 15\, x - 2\, x^2 + 21\,\chi^2\, (1 + x)\,]. \nonumber 
\end{eqnarray}

\section{Matching condition}\label{4}
Since all compact structures are bounded objects, to ensure a well behaved stellar interior i.e finite material content and smooth geometry at the surface $\Sigma\equiv r=r_{s}$ (where $r_{s}$ is the radius of the sphere) of the configuration, one needs to join the inner space-time $\mathcal{M}^{-}$ with the corresponding outer space-time $\mathcal{M}^{+}$. In this case we are trying with uncharged anisotropic fluid sphere described by (\ref{garv1}) and (\ref{Sol1}). Moreover, due to the election given by (\ref{frt}) on the f(R,$\mathcal{T}$) function, the appropriated exterior space-time $\mathcal{M}^{+}$ corresponds to Schwarzschild geometry given by   
 \begin{eqnarray}\label{4.1}
 ds^{2} = -\left(1-\frac{2M}{r}\right)^{-1}\,dr^{2}-r^{2} (d\theta ^{2}+\sin ^{2} \theta  d\phi ^{2} )\nonumber\\+\left(1-\frac{2M}{r}\right) dt^{2},~~~~
\end{eqnarray}
this is so because the modification introduced in matter sector represented by $\mathcal{T}$ is vanishing beyond $\Sigma$. So, to join the interior geometry with Schwarzschild outer space-time one requires to impose the so called first and second fundamental forms across $\Sigma$. The first fundamental form refers the continuity of the intrinsic metric $g_{\mu\nu}$ induced by both metrics $\mathcal{M}^{-}$ and $\mathcal{M}^{+}$ on $\Sigma$. Explicitly it reads 
\begin{equation}
\left[ds^{2}\right]_{\Sigma}=0 \Rightarrow e^{\lambda^{-}(r_{s})}=e^{\lambda^{+}(r_{s})} \quad \mbox{and} \quad e^{\nu^{-}(r_{s})}=e^{\nu^{+}(r_{s})}.   
\end{equation}
So, for the present model we have
 \begin{eqnarray}\label{4.2}
1- \frac{8\,x_s\,(3+x_s)}{7 (1+\chi+x_s)^2}&=&1-{\frac {2M}{r_s}},\\
 \label{4.3} \Psi^2(r_s)&=& \left( 1-{\frac {2M}{r_s}} \right).
 \end{eqnarray}
The second fundamental form is related with the continuity of the extrinsic curvature $K_{\mu\nu}$ induced by $\mathcal{M}^{-}$ and $\mathcal{M}^{+}$ on $\Sigma$. The continuity of $K_{rr}$ component across $\Sigma$ yields to 
\begin{equation}\label{4.4}
\tilde{p}_{r}(r_{s})=0.    
\end{equation}
The above requirement determines the size of the object \i.e the radius $r_{s}$ which means that the material content is confined within the region $0\leq r\leq r_{s}$. The continuity of the remaining components $K_{\theta\theta}$ and $K_{\phi\phi}$ leads to
\begin{equation}\label{eq38}
\tilde{m}(r_{s})=M.   
\end{equation}
Equation (\ref{eq38}) is the total effective mass contained in the sphere which is expressed by Eq. (\ref{eq20}). 
After solving Eqs.(\ref{4.2})-(\ref{4.4}) we obtain the parameters $C$, $D$ and $A$ as,
\begin{eqnarray}\label{DC}
\frac{D}{C}&=&\frac{2 (1 + \alpha x_s)^3\,\sqrt{\Psi_2(x_s)}\, \big[3 + x_s-\alpha\,\Psi_7(x_s)\,\big]}{\big[F_ 1 (x_s) + x^2_s\, F_ 2 (x_s) + x^3_s \, F_ 3 (x_s) + F_{11}(x_s)\big]},~~~~~\\ \label{C1}
C&=&\frac{1}{\Psi_1(x_s)\,\big[1+\frac{D}{C}\,F(x_s)\big]}\sqrt{\frac{7-10\,x_s-x^2_s}{7(1+\chi+x_s)^2}} \\ \label{C2}
A&=&\frac{6r_s-7\,(1 + \chi)\,M \pm 2 \sqrt{7r_s\,(\chi^2- \chi-2) M + 9 r^2_s}}{(7 M - 4 r_s)\, r^2_s},~~~~~
\end{eqnarray} 
where, $x_s=Ar^2_s$,\\$\Psi_7(x_s)=\big(7+7 \,\chi^2-13 \, x_s-2\,x^2_s+14\,\chi\,(1 + x_s)\big)$,\\$ F_{11}(x_s)=x _s\, F_ 4 (x_s) - 7 \, \chi^2 \, F_ 6 (x_s) - \chi \, \big[F_ 7 (x_s) + x^2_s \, F_ 5 (x_s)\big]$. The expressions for the coefficients used here are mentioned in the Appendix \ref{A}. 
Then, Eqs. (\ref{DC})-(\ref{C2}) obtained from Israel-Darmois \cite{r47,r48} junction conditions (first and second fundamental forms) are the necessary conditions to get the compete constants parameter that characterize the model.

\section{geometric characterization of the model}\label{5}
\begin{figure}[htp!]
\includegraphics[width=7.5cm]{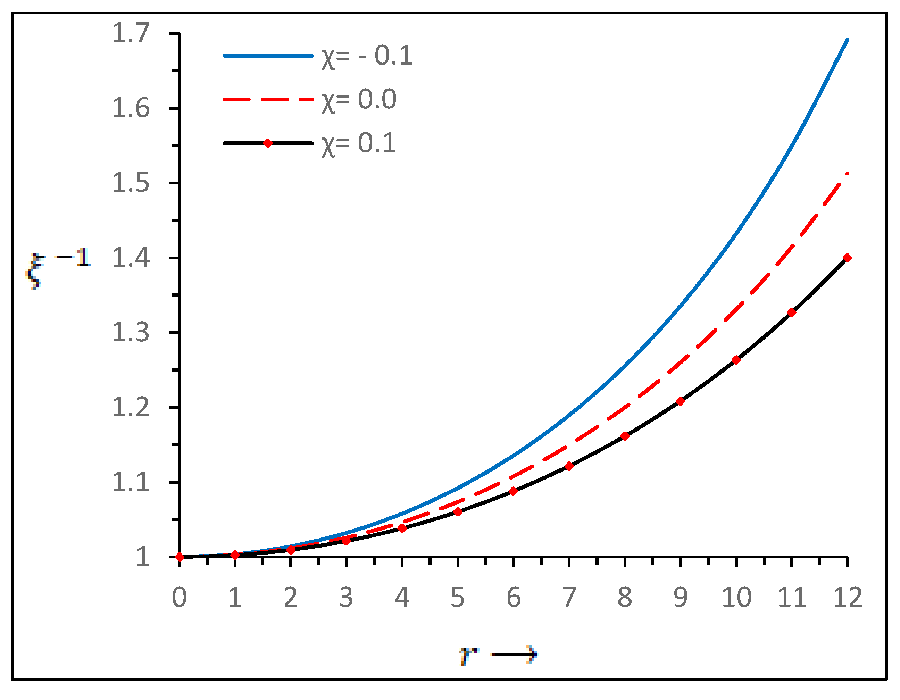} \includegraphics[width=7.5cm]{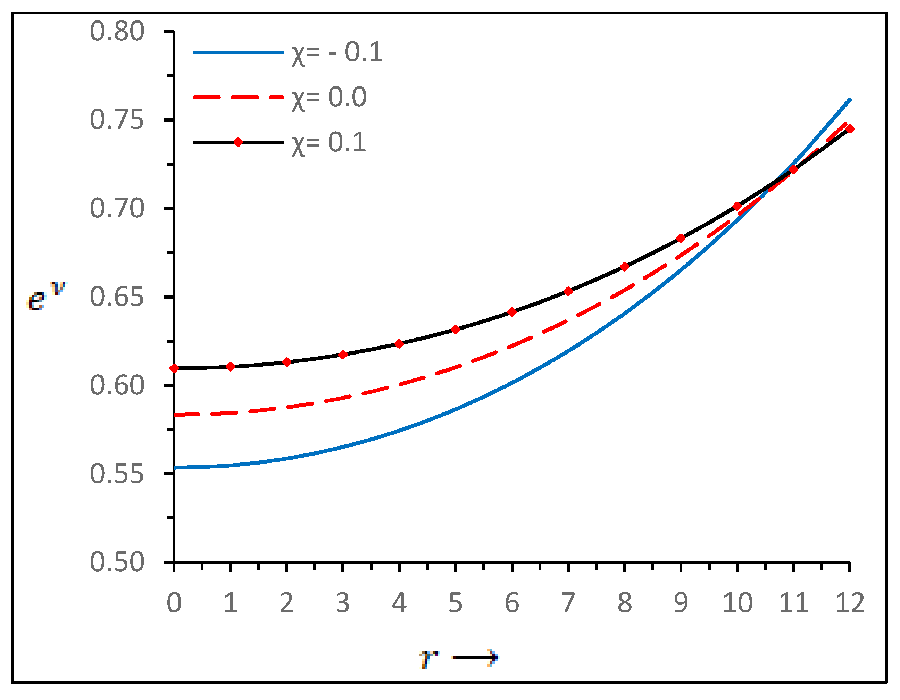}
\caption{ \label{metricpot} Metric potentials $\xi$ and $e^{\nu}$ versus radial coordinate $r$ for f(R,$\mathcal{T}$) gravity theory and Einstein general relativity.}
\end{figure}

The studied model is described by a spherically symmetric static manifold whose temporal and radial metric tensor components are given by Eqs. (\ref{garv1}) and (\ref{Sol1}), respectively. These metric potentials are free from physical and mathematical singularities throughout the compact object. This fact guarantees a well behaved space-time region.
Once the parameters $A$, $C$ and $D$ have been determined, the behavior of the metric potentials is studied through a graphical analysis. As shown in Fig. \ref{metricpot} both $\xi^{-1}$ and $e^{\nu}$ are regular function with increasing radial coordinate everywhere inside the star for all values of the parameter $\chi$. As usual $\xi^{-1}$ takes the value $1$ at $r=0$ while $e^{\nu}>0$ at the same point. The choice of Durgapal-Fuloria type potential $\xi$ has well-founded physical reasons, it is clearly free from physical and mathematical discontinuities.  The extension carried out in this work to the context of f(R,$\mathcal{T}$) gravity theory maintains the same spirit. However, we have made a small modification in order to include the effects of this modified gravitational theory in the full obtained model. From Fig. \ref{metricpot} the dashed red line ($\chi=0.0$) represents the anisotropic Durgapal-Fuloria model in the GR framework. Moreover, the effect of $\chi$ on both metric potentials is clear. Taking $\chi=-0.1$ the potential $e^{\nu}$ at the center is smaller than when $\chi=0.1$ but for $\xi^{-1}$ we have the reverse situation \i.e for $\chi=-0.1$ the curve takes larger values towards the boundary $\Sigma$ than $\chi=0.1$. On the other hand, GR curve behaviour is an intermediate value between them.
\section{Thermodynamic observables }\label{6}
\begin{figure*}
\includegraphics[width=7.5cm]{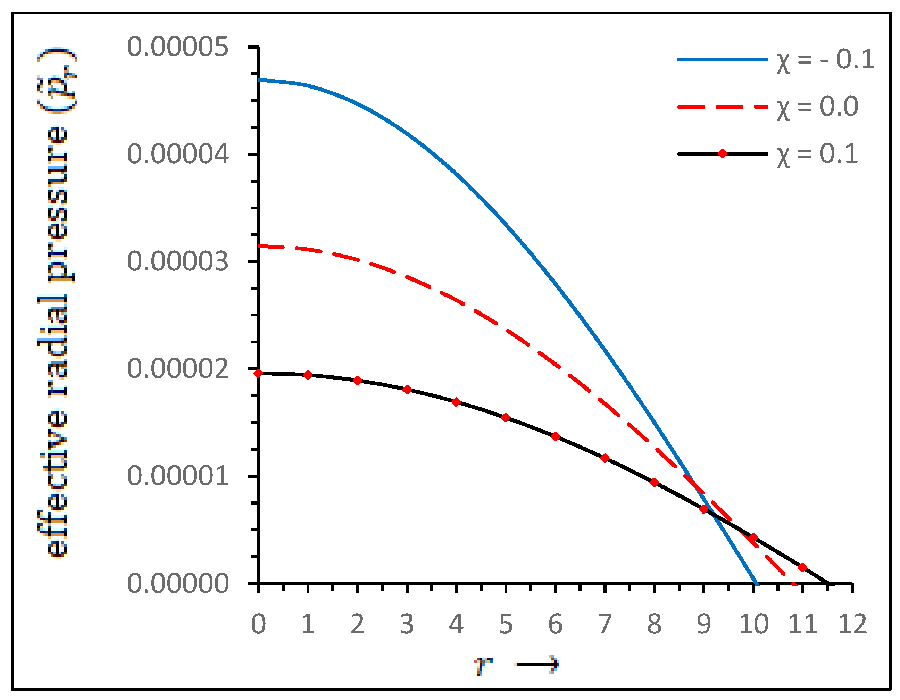}~~ \includegraphics[width=7.5cm]{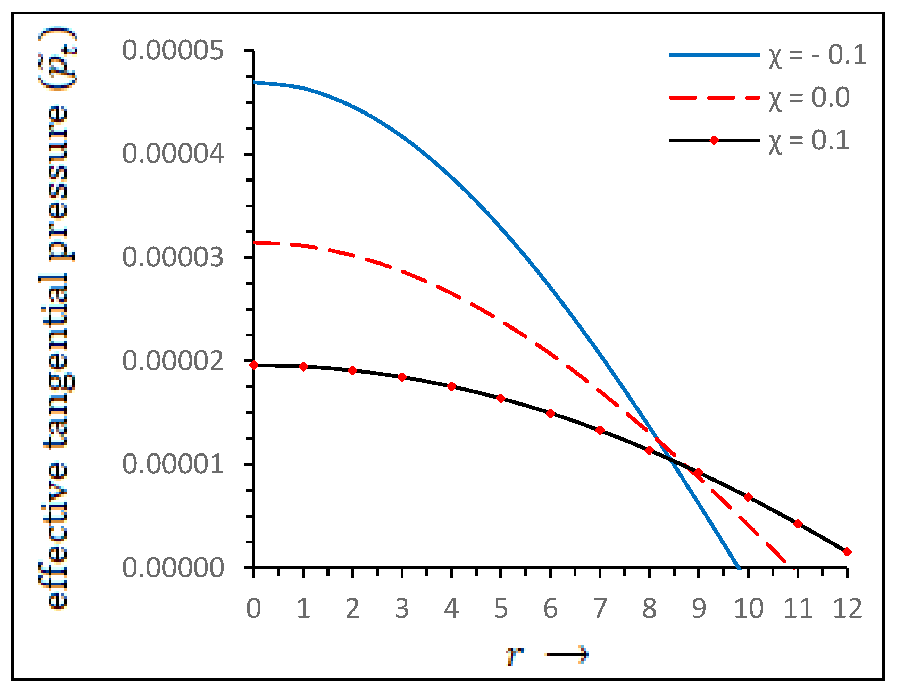}\\
\includegraphics[width=7.5cm]{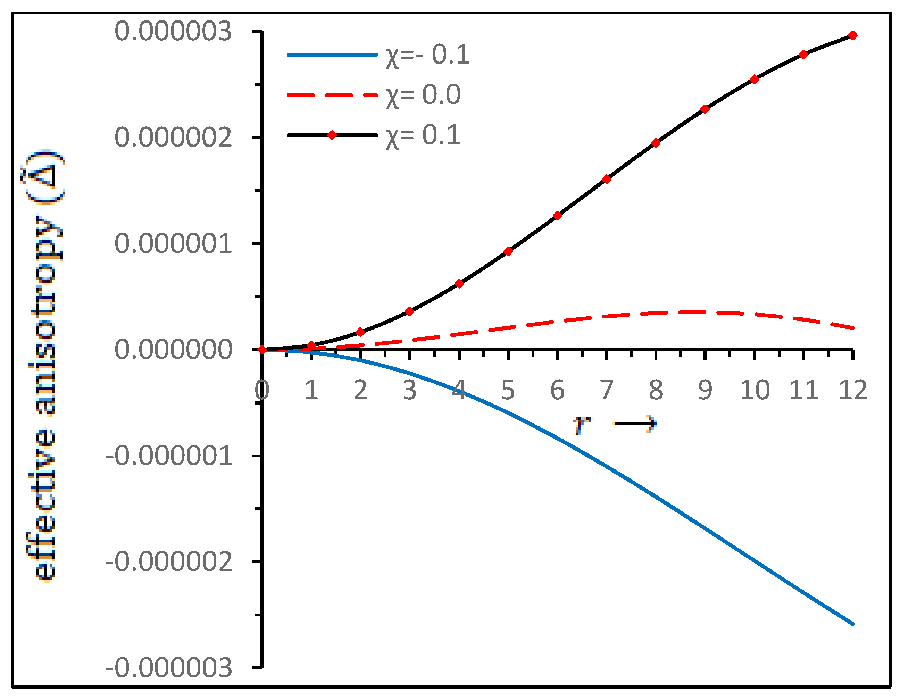}~~ 
\includegraphics[width=7.5cm]{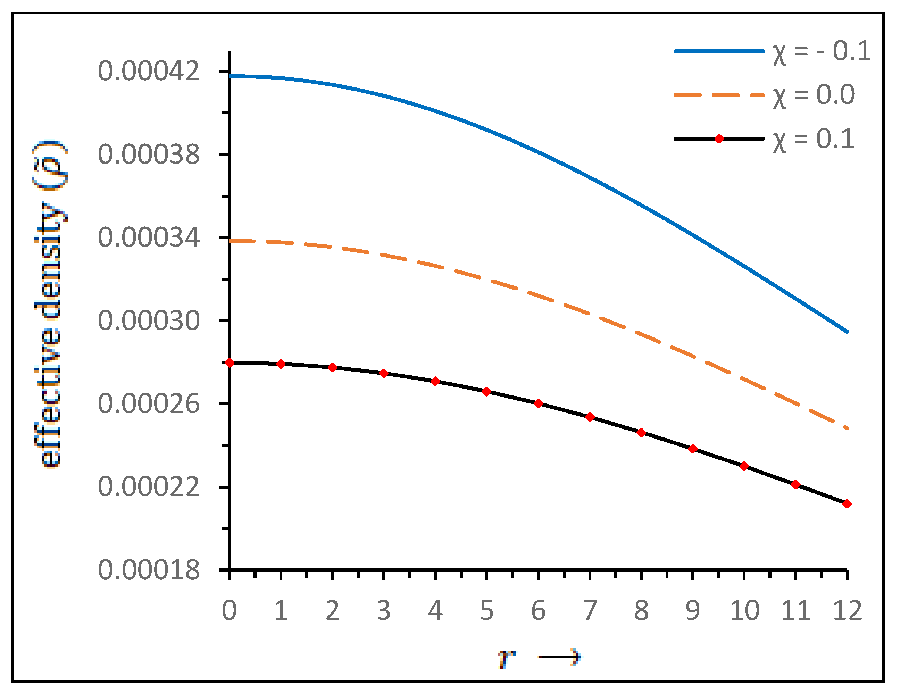}
\caption{\label{fig1} Effective radial pressure ($\tilde{p}_r$), effective tangential pressure ($\tilde{p}_t$), effective energy density ($\tilde{\rho}$) and effective anisotropy ($\tilde{\Delta}$) corresponding to $f(R,\mathcal{T})$ gravity theory (Solid lines) and Einstein general relativity (Dash lines) verses radial coordinate $r$. We have plotted the fig.1 for $A=8.2708 \times 10^{-4}$, $Mass=1.04 M_{\odot}$ for $\chi= -0.1,~0.0,~0.1$. We observe that   effective central pressures and effective central density increases when $\chi$ move form $-0.1$ to $0.1$. We note that larger sphere admits under the $f(R, \mathcal{T})$ gravity theory as compared to Einstein general relativity. We also note that the effective anisotropy is positive and increasing for $\chi=0.1$ and negative decreasing for $\chi=-0.1$ which implies that anisotropic force for $f(R, \mathcal{T})$ gravity is directed outward if $\chi=0.1$ and  directed inward if $\chi=-0.1$. On the other hand for Einstein general relativity, the anisotropy is positive throughout and attains its maximum value within the stellar compact objects. We also note that the configuration is more compact in f(R,$\mathcal{T}$) gravity as compared to Einstein general relativity for $\chi\ge0$. }
\end{figure*}

In this section we study and analyze the behaviour of the main salient features of the model \i.e matter density $\tilde{\rho}$, radial $\tilde{p}_{r}$ and tangential $\tilde{p}_{t}$ pressure respectively. Also we examine the role played by the effective anisotropy factor $\tilde{\Delta}$ inside the stellar structure. \\
It is well known that the main physical parameters of any compact object describing stellar interiors should be free from physical and mathematical drawbacks. Furthermore, they should be  monotonic decreasing functions of the radial coordinate towards the surface, with their maximum values attached at the center of the configuration. These general requirements ensure in principle a well behaved model which can serves to describe some natural objects like white dwarf, neutron stars even quark stars. Moreover, in the study of compact structures there are other ingredients as essential as the aforementioned, which provide a more realistic view of the behavior of celestial bodies. For example, the inclusion of anisotropies in the material content contained in the fluid sphere. Anisotropy in this context means that the pressure in the radial direction is different from the pressure in the angular directions \i.e $p_{r}\neq p_{t}$. So, the effective anisotropy factor is defined by $\tilde{\Delta}=\tilde{p}_{t}-\tilde{p}_{r}$. The inclusion of anisotropies within the stellar content introduces improvements in stability and balance mechanisms and increases the value of the surface redshift. However, regarding the equilibrium mechanism the contribution that it will give depends on the sign, that is, whether it is positive $\tilde{\Delta}>0\Rightarrow \tilde{p}_{t}>\tilde{p}_{r}$ or negative $\tilde{\Delta}<0\Rightarrow \tilde{p}_{t} < \tilde{p}_{r}$. In the first case the system experiences a repulsive force that helps to counteract the gravitational gradient and in the second case, the force due to the anisotropy helps the gravitational force compress the object. If the pressure exerted by the nuclear force fails to overcome the gravitational attraction, the structure eventually will continue to collapse until its Schwarzschild radius. At this point, the object forms a black hole with many unusual properties.  This means that the presence of an attractive force due to anisotropies damages the balance and stability of the configuration. It is clear that the collapse of the structure towards a singularity depends on the gradient pressure (hydrostatic force) exerted by the matter inside the star.
Figure \ref{fig1} shows the behaviour of all thermodynamic observables and anisotropy factor. From the upper panels we can see the behaviour of both effective radial and tangential pressure (left and right respectively). These physical quantities have their maximum values at the center of the configuration and are monotonic decreasing functions with increasing radial coordinate. It is observed that for negative values of $\chi$ the maximum value is greater than the values obtained considering $0.0$ (GR limit) and $0.1$. Respect to the effective density (lower right panel), it has its maximum value attained at the center corresponding with $\chi=-0.1$, is monotonic decreasing function towards the surface and positive defined everywhere within the star. Then, all the thermodynamic observables increases at the center of the star when $\chi$ moves from $-0.1$ to $0.1$. The behavior of the effective anisotropy factor $\tilde{\Delta}$ (lower left panel) is strongly dependent on the value that $\chi$ takes. For $\chi=0.1$ its behaviour is positive at all points within the star, vanishing at the center and increasing function with increasing radius. As we explained above, this conduct introduces in the system a repulsive force (outward). On the other hand, for $\chi=-0.1$ the system is subject to an attractive force (inward). The effect on the system caused by this attractive force will be analyzed in the dynamical equilibrium section \ref{5.2}. Finally, the GR limit corresponding to $\chi=0.0$ shows a positive anisotropy factor throughout the object and attains its maximum value within the configuration. Furthermore, comparing GR ($\chi=0.0$) with $\chi<0$ it is observed that in f(R,$\mathcal{T}$) gravity the object are more compact than in GR. Table \ref{table1} displays the effect of $\chi$ on various physical parameters of the star such as the radius $r_{s}$, central and surface effective energy density, central effective pressure, surface redshift $Z_{s}$ and compactness factor $u=M/R$ for the same Mass$=1.04M_{\odot}$. As we can see, moving $\chi$ from $-0.1$ to $0.1$ all these quantities (except $r_{s}$ and $Z_{s}$) are decreasing in magnitude.

\section{energy conditions}\label{ener}
\begin{figure}[htp!]
\includegraphics[width=8cm]{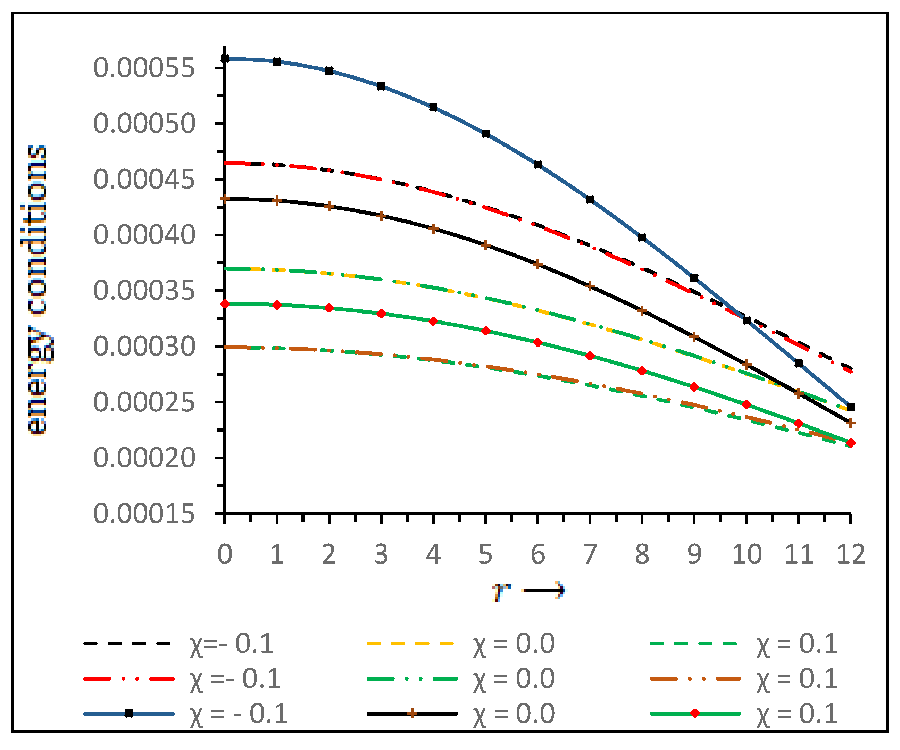} 
\caption{\label{EC} Behavior of energy conditions versus radial coordinate $r$ for $f(R,\mathcal{T})$ gravity theory and Einstein general relativity.In this figure, we have plotted i). small dash lines for $\rho+p_r$, ii). long-dash with dotted lines for $\rho+p_t$, and iii). solid lines for $\rho+p_r+2\,p_t$.}
\end{figure}

It is well known that the matter distribution that makes up celestial bodies can be composed of a large number of material fields. Despite knowing the components that describe this material content inside the compact structure, it could be very complex to describe exactly the shape of the energy-momentum tensor. In fact, one has some ideas on the behaviour of the matter under extreme conditions of density and pressure.  \\
On the other hand, there are certain inequalities which are physically reasonable to assume for the energy-momentum tensor. So, in this section we are willing to verify these inequalities at all points in the interior of the star. In the literature these inequalities are known as energy conditions. Then we have the null energy condition (NEC), dominant energy condition (DEC), strong energy condition (SEC) and weak energy condition(WEC). Explicitly, these are given by 
\begin{eqnarray}
\text{WEC} &:& T_{\mu \nu}l^\mu l^\nu \ge 0~\mbox{or}~\rho \geq  0,~\rho+p_i \ge 0 \\
\text{NEC} &:& T_{\mu \nu}t^\mu t^\nu \ge 0~\mbox{or}~ \rho+p_i \geq  0\\
\text{DEC} &:& T_{\mu \nu}l^\mu l^\nu \ge 0 ~\mbox{or}~ \rho \ge |p_i|\\
&& \mbox{where}~~T_{\mu \nu}l^{\mu} \in \mbox{nonspace-like vector} \nonumber \\
\text{SEC} &:& T_{\mu \nu}l^\mu l^\nu - {1 \over 2} T^\lambda_\lambda l^\sigma l_\sigma \ge 0 ~\mbox{or}~ \rho+\sum_i p_i \ge 0. 
\end{eqnarray}
where $i\equiv (radial~r, transverse ~t),~l^\mu$ and $t^\mu$ are time-like vector and null vector respectively. \\ 
To verify a well defined energy-momentum tensor everywhere within the compact configuration the above inequalities must satisfy simultaneously.
We will check the energy conditions with the help of graphical representation. In Fig. \ref{EC}, we have plotted the L.H.S of the above inequalities which verifies that  all the energy conditions are satisfied at the stellar interior.\\
These energy conditions beyond capturing the idea that the energy must be positive defined, have a clear physical and geometric interpretation \cite{curiel}. From the physical point of view NEC means that an observer traversing a null curve will measure the ambient (ordinary) energy density to be positive. WEC implies that the energy density  measured by an observer crossing a timelike curve is never negative. SEC purports that the trace of the tidal tensor measured by the corresponding observers is always non-negative and finally DEC stand for mass-energy can never be observed to be flowing faster than light. Furthermore, violations of the energy conditions have sometimes been presented as only being produced by unphysical stress energy tensors. Usually SEC is used as a fundamental guide will be extremely idealistic. Nevertheless, SEC is violated in many cases, e.g. minimally coupled scalar field and curvature-coupled scalar field theories. It may or may not imply the
violation of the more basic energy conditions i.e. NEC and WEC. It is worth mentioning that both SEC and DEC imply NEC and DEC implies also WEC. Additionally, the fulfillment of SEC and DEC conditions imposes strong restrictions on the maximum plausible bound of the surface redshift $Z_{s}$ of the compact structure when there are anisotropies in the stellar interior. These implications will be discussed in more details in the next section. 

\section{Causality and surface redshift}\label{7}
\begin{figure}[htp!]
\includegraphics[width=7.5cm]{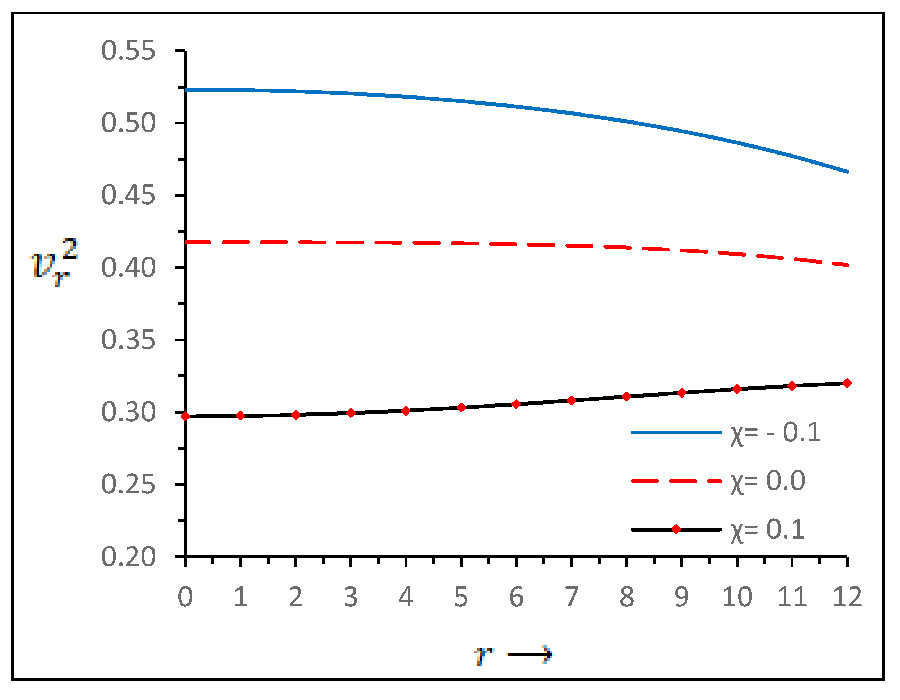} \\
\includegraphics[width=7.5cm]{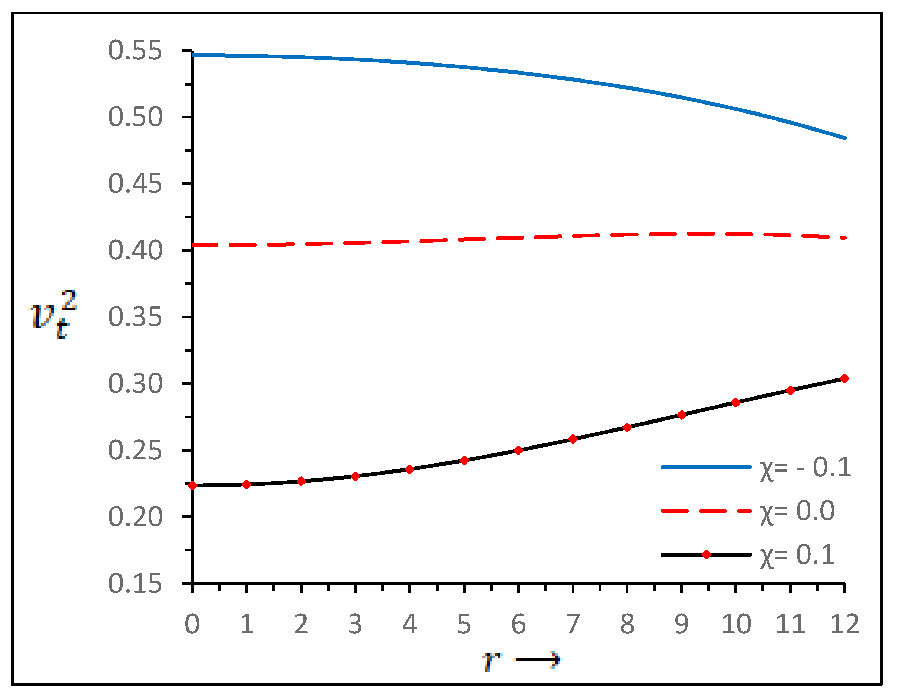}
\caption{\label{V} Behavior of velocity of sounds verses radial coordinate $r$ for $f(R, \mathcal{T})$ gravity theory and Einstein general relativity. 
From the Fig.\ref{V}, we observe that both effective velocities $v^2_r$ and $v^2_t$ are satisfying the causality condition i.e. $0<v^2_r<1$ and $0<v^2_t<1$ everywhere within the stellar models. In $f(R,\mathcal{T})$ gravity system, the radial velocity of sound is greater than the tangential velocity for $\chi=0.1$ while radial velocity of sound is less than the tangential velocity for $\chi=-0.1$. But in scenario of Einstein general relativity the radial velocity of sound is greater than the tangential velocity for $0\le r \le 8.802$ and then start opposite behavior for $r>8.802$.} 
\end{figure}

\begin{figure}[h]
\includegraphics[width=8cm]{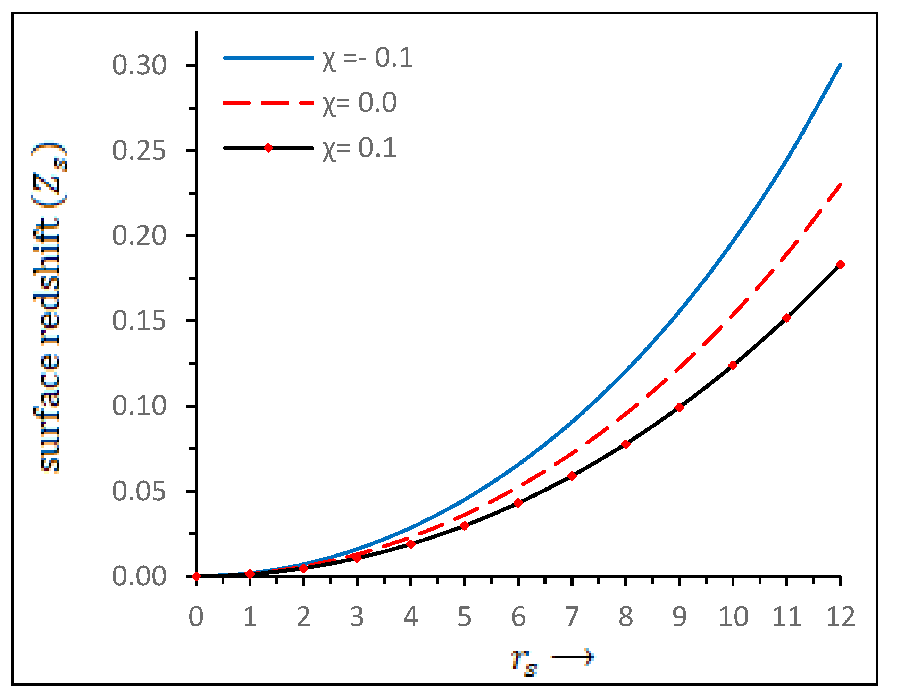} 
\caption{\label{Z} Redshift (Z) verses radius ($r_s$) for $f(R,\mathcal{T})$ gravity theory and Einstein general relativity. From this figure we note that surface redshift is increasing away from centre of stellar objects. It is interesting to see that the surface redshift has more value in $f(R,\mathcal{T}$ gravity theory as compared to Einstein general relativity for $\chi\ge0$.} 
\end{figure}

Among the modifications introduced by the presence of anisotropies in the stellar interior. We have the velocities of propagation associated with the pressure waves in the main directions of the sphere, that is, in the radial and transverse directions and the modification of the upper bound of the surface redshift $Z_{s}$. First of all, the subliminal speeds corresponding to each direction are defined by
\begin{equation}
v^{2}_{r}=\frac{d\tilde{p}_{r}}{d\tilde{\rho}} \quad \mbox{and} \quad v^{2}_{t}=\frac{d\tilde{p}_{t}}{d\tilde{\rho}}.   
\end{equation}
In order to obtain a physically admissible model, both speeds $v_{r}$ and $v_{t}$ must be bounded  by the speed of light ($c=1$ in relativistic geometrized units). This tells us that the pressure (sound) waves in the fluid do not propagate at arbitrary speeds. This is known as causality condition. This condition is peremptory regardless if the material content of the star is isotropic or anisotropic. The only difference between the mentioned cases is that for the anisotropic case there is propagation in the two main directions of the sphere \i.e radial and transverse directions. Moreover, in the isotropic case the subliminal sound speed, should be a decreasing function. However, this is not true in the case where there is anisotropy, since the  speed behaviour depends on the rigidity of the material. So, causality condition reads
\begin{equation}\label{eq43}
0\leq v_{r}\leq 1 \quad \mbox{and}\quad  0\leq v_{t}\leq 1.  
\end{equation}
Causality condition (\ref{eq43}) has strong implications on the behavior of the matter distribution within the object. One of them is related to energy-momentum tensor that describes the material content. If causality is preserved then the energy-momentum tensor is well defined. Secondly, imposing this important condition and once the radial pressure $\tilde{p}_{r}$ is obtained, a relation between it and the density $\tilde{\rho}$ can be established \i.e an equation of state $\tilde{p}_{r}=\tilde{p}_{r}(\tilde{\rho})$. This last statement is very important because usually in searching  solutions to Einstein field equations, the equation of state is imposed. However, it often results in the violation of causality. Additionally, the fact of having different speeds in the directions mentioned above, influences the stability of the system (this subject will be discussed in section A). From Fig. \ref{V} it is appreciated that $v^{2}_{r}$ and $v^{2}_{t}$ are satisfying causality condition for all $\chi$. For $\chi<0$ both subliminal sound speeds are decreasing in nature. Nevertheless, $v^{2}_{t}$ is greater than $v^{2}_{r}$ at all points in the star. In the case $\chi>0$ the radial speed is always greater than the tangential one, whilst in GR scenario the radial sound velocity is greater than the tangential velocity for $0<r<8.802$ and then has an opposite behavior for $r>8.802$.\\
The surface redshift $Z_{s}$ a significant observational parameter that relates the mass $\tilde{m}(r_{s})=M$ and the radius $r_{s}$ of the star, is affected when anisotropies are introduced into the system, regardless the mechanism that originated them. For isotropic fluid spheres its maximum value is $Z_{s}=2$. This value is determined by Buchdahl constraint on the compactness factor $u={2M}/{r_{s}}\leq 8/9$ \cite{r51}. The explicit relation between $Z_{s}$, $M$ and $r_{s}$ is given by
\begin{equation}
Z_{s}=\left(1-\frac{2M}{r_{s}}\right)^{-1/2}-1.
\end{equation}
So, the effect of anisotropies on $Z{s}$ has a long history. For example, Bowers and Liang \cite{r29} considered an hypothetical model containing a constant density $\rho=\rho_{0}$ (incompressible fluid) and a specific form of the anisotropy factor $\Delta$. They concluded that when the anisotropy factor is null i.e $\Delta=0 \Rightarrow p_{r}=p_{t}$ the maximum value for the surface redshift corresponds to $Z_{s}=4.77$, and in the case of a positive anisotropy factor $\Delta>0 \Rightarrow p_{t}>p_{r}$ the above value can be exceed (otherwise if $\Delta<0$). Moreover, if the anisotropy factor is extremely large then the surface redshift will be too.
Moreover, Ivanov studies shown that for realistic anisotropic star models obeying SEC the maximum surface redshift is $Z_{s}=3.842$ (this value corresponds to a model without cosmological constant) meanwhile for models satisfying DEC is given by $Z_{s}=5.211$ \cite{r45}. These values correspond to the following mass-radius relation $0.957$ and $0.974$, respectively. 
In Fig. \ref{Z} the surface redshift $Z_{s}$ has a monotonic increasing behaviour towards the boundary with its maximum value attained at the boundary of the object. Besides, for negative values of $\chi$ the surface redshift takes larger values in comparison with $\chi\geq0$. This is also appreciated in table \ref{table1}. Although Ivanov's research reveals that the value of $Z_{s}$ in the presence of anisotropic fluids exceeds the Buchdahl bound, the values reported for $Z_{s}$ in this work are below the maximum values reported for both isotropic and anisotropic distributions, but are in complete agreement with what has already been established, in the sense that they do not exceed such values.

\section{Dynamical Equilibrium condition of stellar model in $f\left(R,\mathcal{T}\right)$ theory gravity} \label{5.2}
\begin{figure}[htp!]
\includegraphics[width=8cm]{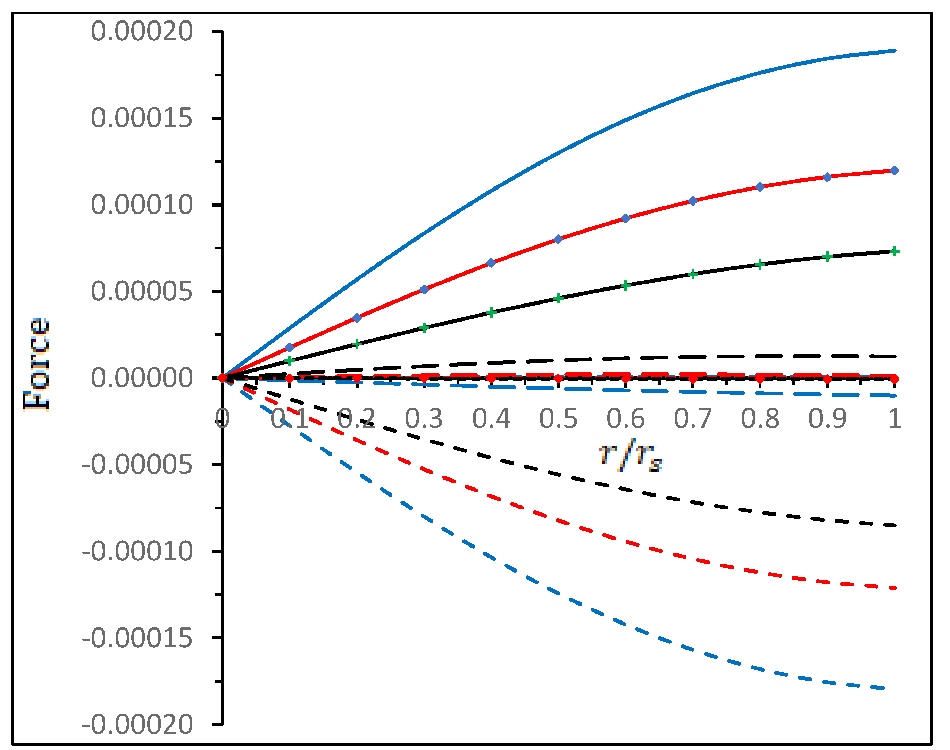} 
\caption{\label{force} Variation of different forces hydrostatic force $F_h$ (solid lines), gravitational force $F_g$ (small dash lines), anisotropic force $F_a$ (long dash lines), coupling force $F_{\chi}$ (dotted lines) verses radial coordinate $r/r_s$. For this figure we have used the following values of parameters:(i) $\chi= -0.1$, $r_s=10.062$ for blue lines, (ii) $\chi=0,~~r_s=10.801$ for red lines, (iii) $\chi=0.1,~~r_s=11.517$ for black lines. From this figure we note that the hydrostatic force $F_h$ for $f(R, \mathcal{T})$ has more value than the hydrostatic force $F_h$ for Einstein GR if $\chi< 0$ and reverse behaviour if $\chi>0$. Moreover, The anisotropic force $F_a$ is inward if $\chi=- 0.1$ and outward if $\chi = 0.0 $ and 0.1. The coupling force $F_{\chi}$ has very less effect to balance this mechanism.} 
\end{figure}

 In this section we discuss the dynamical equilibrium condition of the stellar model by using Tolman-Oppenheimer-Volkoff  (TOV) \cite{r49,r50} equation in the framework of $f\left(R,\mathcal{T}\right)$ theory gravity. This modified TOV equation for $f\left(R,\mathcal{T}\right)$ theory, as already mentioned by Eq.(\ref{TOV}), is given by 
  \begin{eqnarray}\label{eq42}
 -\frac{dp_r}{dr}-\frac{\nu^{{\prime}}}{2} \big( \rho+&&p_r \big)+\frac{2}{r}\left({p_t}-{p_r}\right)\nonumber\\+&&{\frac {\chi}{6(4\pi+\chi)}}\left(3\,\frac{d\rho}{dr}-\frac{dp_r}{dr}-2\,\frac{dp_t}{dr} \right)=0.~~~~~~
\end{eqnarray}
where we denote first term $-\frac{dp_r}{dr}=F_h$, second term $-\frac{\nu^{{\prime}}}{2} \big( \rho+p_r \big)=F_g$, third term $\frac{2}{r}\left({p_t}-{p_r}\right)=F_a$ and fourth term ${\frac {\chi}{6(4\pi+\chi)}}\left(3\,\frac{d\rho}{dr}-\frac{dp_r}{dr}-2\,\frac{dp_t}{dr} \right)=F_{\chi}$. These term describe the hydrostatic force ($F_h$), gravitational force ($F_g$), anisotropic force ($F_a$) and coupling force ($F_{\chi})$, respectively. \\
In the case of isotropic fluid spheres ($\tilde{p}_{r}=\tilde{p}_{t}$) and regarding $\chi=0.0$ (GR limit), this equation drives the equilibrium of relativistic compact structures such as neutron stars, white dwarfs, etc. In considering the inclusion of anisotropies and the effect of relativistic modified gravity theories, this equation still drives the balance of the system. However, its form change a little bit as shown Eq. (\ref{eq42}). As we can see this equation relates the effective thermodynamic quantities with the metric potential $e^{\nu}$. In order to keep the system in equilibrium and prevent it from collapsing below its Schwarzschild radius, it is necessary to have a relationship between the thermodynamic variables $\tilde{p}_{r}$ and $\tilde{\rho}$, that is, an equation of state (EoS) $\tilde{p}_{r}=\tilde{p}_{r}(\tilde{\rho})$ that links them. Nonetheless, in this case where there are contributions from the anisotropies and from the theory considered, for certain conditions it may not be enough to withstand the gravitational attraction. Thus
the structure equations (\ref{eq20}) and (\ref{eq42}) imply that there is a maximum mass that a star can have. As was pointed out before, the present model is under four forces. The impact that these have on the system is shown in fig. \ref{force}. We can observe that the system is in equilibrium, the gravitational gradient is counterbalances by the hydrostatic $F_{h}$, anisotropic $F_{a}$ and coupling $F_{\chi}$ forces (although its contribution is very small) when $\chi>0$. On the other hand, when $\chi<0$ the anisotropic force takes negative values. It means that the system is under an attractive force. This fact his can damage the balance of the system if the hydrostatic gradient is not strong enough to counteract the force exerted by the gravitational attraction and the anisotropic force. Thus the system can collapse towards a singularity. However, as noted, the negative anisotropic force is very small in magnitude and the pressure gradient overcomes the action $F_{g}+F_{a}$.\\
Another interesting point to note is that for positive values of $\chi$ the hydrostatic gradient is lower in f(R,$\mathcal{T}$) theory than in the corresponding one in GR (the inverse situation is presented for $\chi<0$). The same happens with the gravitational attraction.

\begin{table*}
\caption{\label{table1}Comparative study of physical values of the compact star SMC X-1  for $A=8.2708 \times10^{-4}$ and Mass$=1.04M_{\odot}$ $\alpha=1.12$ for different values of $\chi$.}  
\begin{ruledtabular}
\begin{tabular}{ccccccc}
 {$\chi$}& Radius & surface-redshift& Mass-radius ratio & central pressure & central density & surface density \\ 
& $r_s(Km)$&($Z_s$)&$\frac{M}{R}$ & $(\tilde{p}_r)_c$ & $\tilde{\rho}_c$ & $\tilde{\rho}_s$\\ \hline
0.1 & 11.517 & 0.15853 & 0.13319 &$2.38067\times10^{34}$ & $3.77548\times10^{14}$ &  $2.92074\times10^{14}$ \\  
0.0 & 10.801 & 0.15483 & 0.14204  &$3.82116\times10^{34}$ & $4.56834\times10^{14}$ &  $3.54306\times10^{14}$ \\
-0.1 & 10.062 & 0.14597 & 0.15245  &$5.70446\times10^{34}$ & $5.63992\times10^{14}$ &   $4.38951\times10^{14}$ \\  
\end{tabular}
\end{ruledtabular}
\end{table*} 

\section{The Equation of state (EoS)}\label{9}
In the study of compact structures such as neutron stars it is very important to know how the principal thermodynamic variables are connected. This relation known as the equation of state (EoS) drives a relationship between the effective radial pressure $\tilde{p}_{r}$ and the effective energy density $\tilde{\rho}$. The microphysics, as described by the EoS, is linked to the macroscopic properties of the neutron star, in particular, their masses and radii, via the Tolman-Oppenheimer-Volkoff equations, which provide the direct relation that is necessary to use astrophysical observations to constrain nuclear physics at very high densities. However, the composition of a neutron star chiefly depends on the nature of strong interactions. Depending on the type of interaction, the models can be grouped into three broad categories: non-relativistic potential models,
relativistic field theoretical models, and relativistic Dirac-Brueckner-Hartree-Fock models. In addition, in each of these perspectives,  the presence of softening components such as hyperons, Bose condensates or quark matter, can be incorporated \cite{lattimer,prakash}. On the other hand, one can classify the EoS in two classes:  First, normal equations of state have a
pressure which vanishes as the density tends to zero. Second, self-bound equations of state
have a pressure which vanishes at a significant finite density. Respect to the self-bound EoS the most famous example is the MIT bag model EoS. It was pointed out by Witten \cite{witten} that strange quark matter is the ultimate ground state of matter. This leads to the fact that the internal and external vacuum densities of the hadrons are completely different and that the vacuum pressure of the bag wall balances the pressure of the quarks, stabilizing the whole system \cite{alcock,haensel}. So, the MIT bag EoS model reads
\begin{equation}
p=\frac{1}{3}\left(\rho-4B\right),    
\end{equation}
where $B$ is the so called bag constant and represents the difference between the energy density of the perturbative and non-perturbative QCD vacuum. In this model the interactions of quarks and gluons are enough small, neglecting quark masses and supposing that quarks are confined to the bag volume. Concerning normal matter the EoS describes an interacting nucleon gas above a transition
density $1/3\rho_{s}$ to $1/2\rho_{s}$ (being $\rho_{s}$ the surface density). Below this density, the ground state of matter consists of
heavy nuclei in equilibrium with a neutron-rich, low-density gas of nucleons. Nonetheless, the equilibrium of the system exists below the transition density \cite{baym,lattimer1}. So, in order to explain the structural properties of compact stars model at high densities, several authors have proposed the EoS $P=P(\rho)$ should be well approximated by a linear function of the energy density $\rho$
\cite{dey,harko,gondek}. Some authors have also expressed more convincing approximated forms of the EoS $P =P(\rho)$  as linear function of energy density $\rho$ \cite{harko1,mauryaprd2}. Furthermore, a linear relation between pressure $P$ and energy density $\rho$ ensure the preservation of causality condition.\\
For the present model the EoS is not a linear relation between pressure and energy density. The functional relation is more complicated. The effective radial pressure $\tilde{p}_{r}$ in terms of the surface $\tilde{\rho}_{s}$ and central $\tilde{\rho}_{c}$ effective energy density the EoS has the following form
\begin{eqnarray} 
\tilde{p}_r&=&\frac{\tilde{\rho}_c\,\big[-8\,(3+\rho_1)+4\,\Psi_2(\rho_1)\,\Psi_4(\rho_1)\big]\,(1+\chi)^2}{72 \,(1+\chi+\rho_1)^2},\label{51}~~~~~\\
\tilde{p}_{t}&=&\frac{\tilde{\rho}_c\,(1+\chi)^2}{72}\Bigg(\frac{-8\,(3+\rho_1)+4\,\Psi_2(\rho_1)\,\Psi_4(\rho_1)}{(1+\chi+\rho_1)^2}\nonumber\\&& +
\frac{8\,\rho_1\,\big[ \Psi_5(\rho_1)+ \Psi_6(\rho_1)+\chi\,(21 + 6\, \rho_1 + 4\, \rho_1^2)\big]}{(1 + \chi + \rho_1)^3\,(1 + \alpha\,\rho_1)^2}\bigg).~~~~~~ \label{EOSpt}
\end{eqnarray}  
However the expressions for other used coefficients are mentioned in the Appendix \ref{B}. In Fig. \ref{eos} we can graphically appreciate the shape of the EoS of the model under study. In spite of the complex relation given by \ref{51}, the behavior that appears from the surface to the core of the object is linear (the curve grows from regions of low to high densities). This curve can be described approximately by the following polynomial linear interpolation (keeping only the first order in $\tilde{\rho}_{s}$) as follows
\begin{equation}\label{linear}
\tilde{p}_{r}=\alpha\left(\tilde{\rho}-\tilde{\rho}_{s}\right),    
\end{equation}
where $\alpha$ is a non-negative constant. It is clear from Eq. (\ref{linear}) that when $r=r_{s}$ then $\tilde{p}_{r}=0$. This is so because $\tilde{\rho}(r_{s})=\tilde{\rho}_{s}$ \i.e the energy density at zero pressure (surface energy density). Additionally, we can observe from Fig. \ref{eos} the effect of $\chi$ on the EoS. For $\chi=-0.1$ the effective central pressure $\tilde{p}_{r}$, the effective central energy density $\tilde{\rho}$ and the effective surface energy density $\tilde{\rho}_{s}$ are greater than when $\chi\geq 0$. It can also be checked in table \ref{table1}.

\begin{figure}[htp!]
\includegraphics[width=7.5cm]{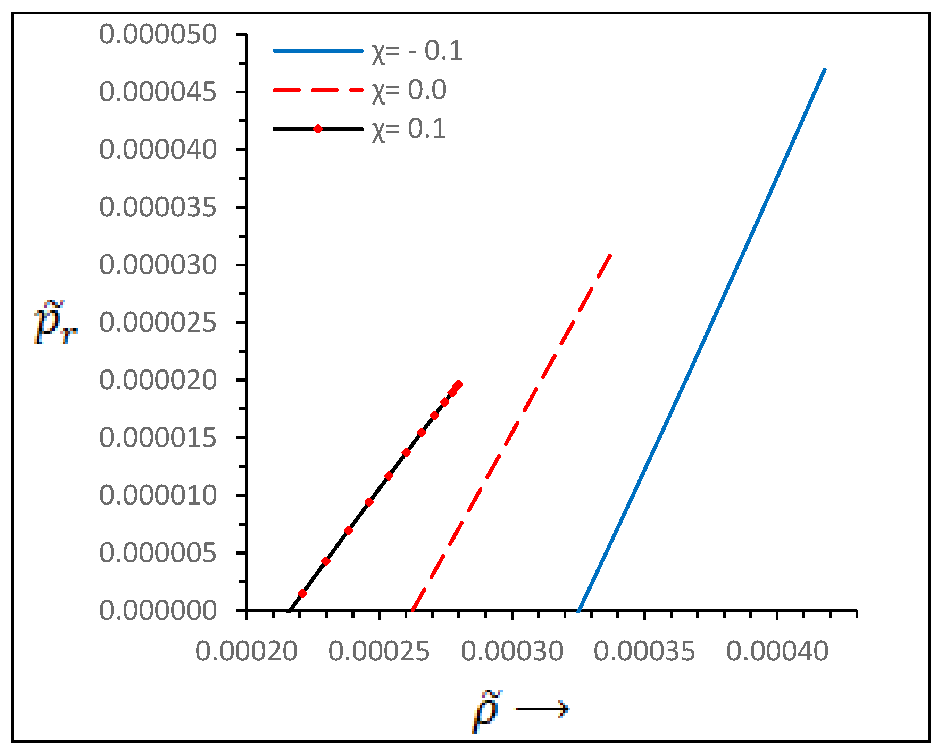} 
\caption{\label{eos} effective pressure($\tilde{p_r}$) verses effective energy density ($\tilde{\rho}$) for $f(R,\mathcal{T})$ gravity theory and Einstein general relativity.}
\end{figure}

\section{Stability}\label{10}
\begin{figure}[htp!]
\includegraphics[width=7.5cm]{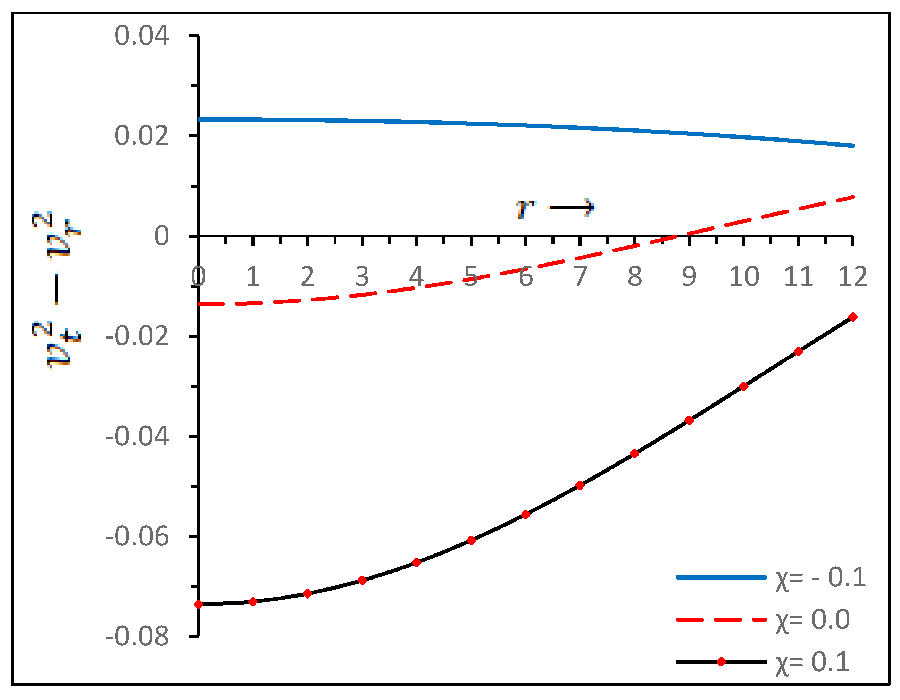}
\includegraphics[width=7.5cm]{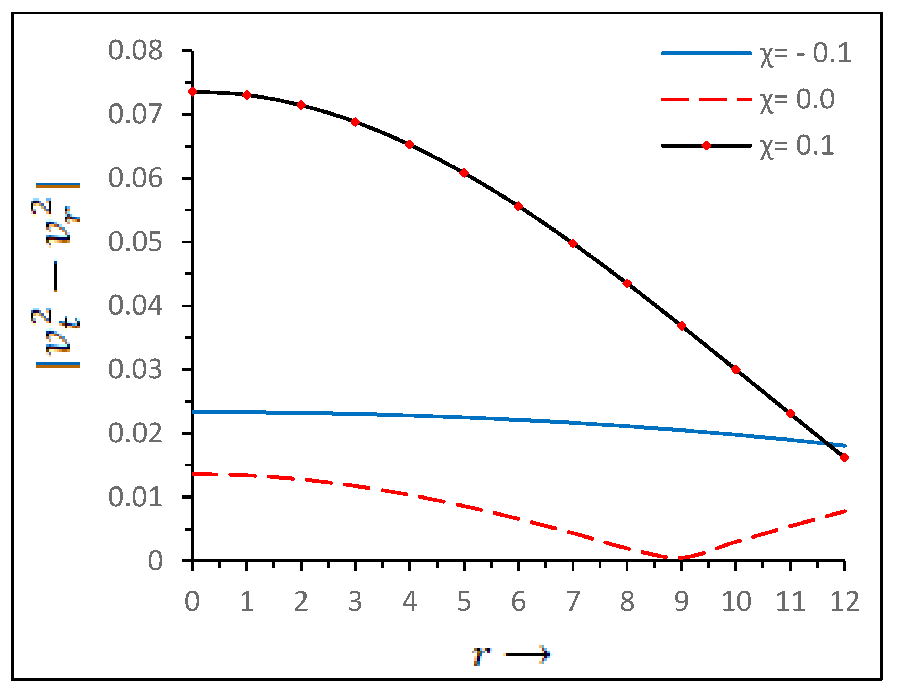}
\caption{\label{abreu} Variation for velocity difference $v^2_t-v^2_r$ and $|v^2_t-v^2_r|$ verses radial coordinate $r$ in framework of $f(R,\mathcal{T})$ gravity and Einstein GR.  From this Fig.(\ref{abreu}), we observe that the velocity difference $v^2_t-v^2_r$ in $f(R,\mathcal{T})$ gravity system is positive for $\chi=0.1$ and negative throughout within the stellar compact star models while in scenario of Einstein general relativity the velocity difference $v^2_t-v^2_r$ is positive for $0\le r \le 8.802$ and negative if $ r> 8.802$. This implies that the cracking does not appear within the anisotropic matter distribution for $f(R, \mathcal{T})$ gravity system while the cracking appears within the anisotropic matter distribution for Einstein general theory of relativity. We conclude that anisotropic compact star model are stable for $f(R, \mathcal{T})$ gravity system as compared to Einstein system.}
\end{figure}

 In this section we analyze the stability of the model by means of Abreu's criterion \cite{r52}. Basically, the method consists in contrasting the speed of pressure waves in the two principal directions of the spherically symmetric star: the subliminal radial sound speed with the subliminal tangential sound speed, and then based on those values at particular points in the object, one could potentially conclude whether the system is stable or unstable under cracking instability. Put forward by Herrera \cite{r37}, cracking involves the possibility of smashing the fluid sphere in view of the appearance of total radial forces of different signs, and therefore in different directions, at different points within the configuration. It should be emphasized that this effect has never been observed, however under appropriate physical assumptions, it is a likely scenario.\\
Cracking process is mechanism to study  instability when anisotropy matter distributions are present. Nevertheless, this mechanism can be characterized most easily through the subliminal speed of pressure waves. Following we have
\begin{equation}
\frac{\delta\tilde{\Delta}}{\delta\tilde{\rho}}\sim \frac{\delta\left(\tilde{p}_{t}-\tilde{p}_{r}\right)}{\delta\tilde{\rho}} \sim \frac{\delta\tilde{p}_{t}}{\delta\tilde{\rho}}-\frac{\delta\tilde{p}_{r}}{\delta\tilde{\rho}}\sim v^{2}_{t}-v^{2}r.    
\end{equation}
Moreover, from causality condition one has $0\leq v^{2}_{r}\leq 1$ and $0\leq v^{2}_{t}\leq 1$ which implies $0\leq |v^{2}_{t}-v^{2}_{r}|\leq 1$. Explicitly it reads
\begin{eqnarray}
     \label{eq46}
   & \quad\hspace{-6.2cm} -1\leq v^{2}_{t}-v^{2}_{r}\leq 1  =\nonumber\\ &
\quad\hspace{0.2cm} \left\{
	       \begin{array}{ll}
		   -1\leq v^{2}_{t}-v^{2}_{r}\leq 0~~ & \mathrm{Potentially\ stable\ }  \\
		 0< v^{2}_{t}-v^{2}_{r}\leq 1 ~~ & \mathrm{Potentially\ unstable}
	       \end{array}
	        \right\}.~~
	    \end{eqnarray}
Therefore, the main idea behind Abreu's criterion is that if the subliminal tangential speed $v^{2}_{t}$ is larger than the subliminal radial speed $v^{2}_{r}$; then this could potentially result in cracking instabilities to occur in the object, rendering the latter an unstable configuration. So, with the help of graphical analysis one can determines the potentially stable/unstable regions within the star and then conclude whether the system is stable or not. From Fig. \ref{abreu} (upper plot) it is observed that the system presents all the regions completely stable, that is, the whole system is stable (for $\chi>0$). Completely unstable behavior (for $\chi<0$), it is also observed that the region $0<r<8.802$ is stable while the region $r>8.802$ is unstable (for $\chi=0$). Then, there is cracking (change in sign of $v^{2}_{t}-v^{2}_{r}$ ) in GR theory for the present model. This is so because from $r>8.802$ the tangential velocity is greater than the radial velocity of the pressure waves. Instability of the system for negative values of $\chi$ could be anticipated from Fig. \ref{V}, since $v^{2}_{t}$ is greater than $v^{2}_{r}$ at every point inside the star. In distinction for positive values of $\chi$ where $v^{2}_{t}$ is always less than $v^{2}_{r}$ everywhere within the object. Although $ |v^{2}_{t}-v^{2}_{r}|$ lies between $0$ and $1$ (lower panel Fig. \ref{abreu}), it does not means that the system is stable. Notwithstanding, it warns us of the presence of cracking in the sphere, like in the case of GR. As seen in the lower panel of Fig. \ref{abreu}, the red curve (GR) is decreasing up to certain value of the radial coordinate $r$ and then suddenly changes its behavior to a growing one. Therefore, in this opportunity for the specific value $\chi=0.0$, the system has stable and unstable regions.

\section{SUMMARY AND OUTLOOK}\label{11}
In the present paper we have obtained an analytic relativistic anisotropic spherical model in the framework of f(R,$\mathcal{T}$) gravity theory. This was achieved through the imposition of three ingredients, which form the fundamental pillars of the obtained model. The first of these was to consider a simple, notable and viable modified gravity model given by $f(R,\mathcal{T})=R+2\chi\mathcal{T}$ \cite{r1}, being $R$ the usual Ricci scalar, $\mathcal{T}$ the trace of the energy momentum-tensor and $\chi$ a constant coupling. In addition we have taken the Lagrangian density matter to be $\mathcal{L}_{m}=\frac{1}{3}\left(p_{r}+2p_{t}\right)$. The second one is the imposition of the metric potential $\xi$, which corresponds to a modification of the original potential proposed by Durgapal-Fuloria \cite{Durgapal} in the context of GR. The modification on $\xi$ considers the inclusion of $\chi$ (see Eq. (\ref{garv1})). Finally, we have imposed the isotropic condition under the restriction $p_{r}\neq p_{t}$ in order to obtain the effective anisotropy factor $\tilde{\Delta}$. With these ingredients in hand we arrive at the differential equation given by Eq. (\ref{Diff2}) in order to get the $e^{\nu}$ metric potential. Once this differential equation is solved the inner geometry of the whole system is completely specified. The election of $\xi$ is well motivated because it is free from physical and mathematical singularities everywhere inside the compact object. Moreover, the complete internal manifold (\ref{garv1})-(\ref{Sol1}) is well behaved at all point within the star. After that we proceeded to obtain the constant parameters of the solution. For this purpose we have made the joint between the model and the external Schwarzschild solution on the surface $\Sigma$ of the compact structure. Thus, the first and second fundamental form provide the corresponding parameter space that characterizes the model. The junction between the collapsed configuration and the Schwarzschild space-time was possible due to the modifications introduced by $\mathcal{T}$ in the matter sector remain finite and bounded by the object. It means that beyond $\Sigma$ we have an empty space-time. \\
It is of our interest to check the outcomes and the predictions of one of the extended gravity, i.e., f(R,$\mathcal{T}$) theory regarding the existence, stability and equilibrium of spherical stars. Therefore, we have explored the behaviour of the main salient features such as the effective radial and tangential pressure, the effective energy density and the effective anisotropy factor. The behaviour of all these quantities is influenced by $\chi$. Throughout the study we have taken $-0.1\leq\chi\leq 0.1$, $M=1.04M_{\odot}$ and  $\alpha=1.12$. For these values the effective thermodynamic observables $\tilde{p}_{r}$, $\tilde{p}_{t}$ and $\tilde{\rho}$ remain positives and their maximum values are attained at the core of the compact star. It means that they have a monotonic decreasing behaviour from the center to the boundary. It is remarkable to note that $\chi=-0.1$ induces greater central values than $\chi=0.1$ and $\chi=0.0$ (GR limit). Respect to the effective anisotropy factor, this quantity has a more intricate behavior, that is, for $\chi=-0.1$ it is  negative decreasing function at every point inside the star, meanwhile for $\chi\geq0$ it is  positive increasing function everywhere within the object. A negative anisotropy factor means that the system is subjected to an attractive force (inward). This fact can damage the equilibrium and stability mechanism. In distinction with the case $\chi\geq0$ where the induced force is completely repulsive (outward) and helps to counteract the gravitational attraction in order to avoid point singularities or the formation of a black hole (the latter occurs when the object is compacted below its Schwarzschild radius). The proper behavior of the thermodynamic variables leads to a well-defined energy-momentum tensor. It is checked employing the so called energy conditions. In this study energy conditions holds for all values of $\chi$. The realization of these conditions has strong implications for the preservation of causality (the material content inside the object does not travel faster than light) and the maximum possible value of the surface redshift, which can not be arbitrarily large when there is anisotropy. Fortunately, for this model the causality is completely preserved and the maximum value that the surface redshift takes for all $\chi$ values considered, does not exceed the maximum level reported in the literature \cite{r45}. The analysis of balance mechanism is studied using the modified version of the TOV
equation in the realm of couple of f(R,$\mathcal{T}$) model. For the particular choice of f(R$\mathcal{T}$) model, it is shown that the system is under four forces, namely the hydrostatic force $F_{h}$, gravitational force $F_{g}$,  anisotropic force $F_{a}$ and the coupling force $F_{\chi}$,
attaining the equilibrium condition by keeping all the forces sum to be zero. It is worth mentioning that the usual GR forces are
being modified due to f(R,$\mathcal{T}$) model, thus producing some
extra effects in the forces $F_{h}$, $F_{g}$ and $F_{a}$. Furthermore, we have arrived at a more compact configuration for $\chi<0$. This is so because, the attractive force due to negative anisotropy helps compact the object in conjunction with the gravitational force. Moreover, we have obtained the corresponding equation of state (EoS) by analytic procedure. This important relation between the effective radial $\tilde{p}_{r}$ and the effective energy density $\tilde{\rho}$ tell us (in some sense) how is the material composition of the star. In addition, the EoS allows to determine the mass-radius relation (it is also related with the surface redshift), \i.e from the microphysic one can get the macro observables of the compact structure. Despite of the complex expression describing the EoS of the model given by Eq. (\ref{51}), its behaviour is approximately linear. This indicates that the interior structure of
the relativistic compact stellar objects is composed of normal ordinary matter. 
The stability of our
compact star depends upon the choices of the  parameter involved in f(R,$\mathcal{T}$) model. The difference of the squares subliminal sound speeds, i.e, $|v^{2}_{t}-v^{2}_{r}|$ has been found to be within
[0, 1], for $\chi=0.0$ the system presents cracking, meanwhile for $\chi=-0.1$ and $\chi=0.1$ we can not conclude anything. To check if the system is stable/unstable we need to study $v^{2}_{t}-v^{2}_{r}$. It was found that the system is completely unstable when $\chi=-0.1$ and completely stable for $\chi=0.1$. Furthermore, for the GR case ($\chi=0.0$) the system has stable and unstable regions (cracking). All the above discussion is supported by an extensive graphic analysis which is states in Figs. \ref{metricpot}, \ref{fig1}, \ref{EC}, \ref{V}, \ref{Z}, \ref{force}, \ref{eos} and \ref{abreu}. Finally, we wish to comment that by taking $\chi=0.0$ and $\alpha=1$ the corresponding results of the Durgapal-Fuloria model reported in \cite{Durgapal} are recovered and that in the case $\chi>0$ the obtained model satisfies all the general requirements in comparison with GR. Then, the existences of compact structures within the framework of f(R,$\mathcal{T}$) is a good opportunity to understand many phenomena in the strong gravitational field regime.

\begin{acknowledgments} 
S. K. Maurya acknowledge continuous support
and encouragement from the administration of University of Nizwa. F. Tello-Ortiz thanks 
the financial support by the CONICYT PFCHA/DOCTORADO-NACIONAL/2019-21190856, grant Fondecyt No. 1161192, Chile and project ANT-1855 at the Universidad de Antofagasta, Chile.
\end{acknowledgments}

\begin{widetext}
\appendix
\section{The expressions for used coefficients in Eqs.(\ref{DC})  and (\ref{C1}) are as follows:} \label{A} 
\begin{eqnarray}
\Psi_ 1 (x_s) &=& (1 + \alpha\, x_s)^2, ~~ ~~ \Psi_ 2 (x_s) = \big[ 7 + 7\chi^2 - 10 x_s - x_s^2 + 14\chi (1 + x_s)\big], ~~ \Psi_ 3 (x_s) = \big[-1 + 7\alpha^2 (1 + \chi)^2 - 2\alpha (-5 + 7\chi)\big], \nonumber\\
F(x_s)&=& \frac{\sqrt{7}\,\sqrt{\Psi_ 2 (x_s)} \big[f_ 1 (x_s) + f_ 2 (x_s)\big]}{3\, \Psi_ 3 (x_s)\,\Psi_1(x_s)\,\sqrt{\Psi_1(x_s)}} +\frac{\sqrt{7}\,f_3 (x_s) }{(\Psi _ 3 (x_s))^{7/2}}\,\ln\bigg[\frac {-a (\Psi_3 (x_s))^{5/2}\, f_ 5 (x_s)} { 2 \, 
     f_ 4 (x_s)\, (1 + \alpha\,x_s)}\bigg],\nonumber\\
f_ 1 (x_s) & = & (1 - \alpha - \alpha\, \chi)\Psi^2 _ 3 (x_s) + \Psi_ 3(x_s)\, \big[-1 + 
    6\,\alpha\,\chi + \alpha^2\, (-23 - 16\chi + 7\chi^2)\big] (1 + \alpha\,x_s), \nonumber\\
f_ 2 (x_s)& = &  \big[1 + \alpha (7 - 23\chi) + \alpha^2 (63 + 174\chi - 105\chi^2) + \alpha^3 (-359 + 201\chi + 399\chi^2 - 161\chi^3)\big]\, (1 + \alpha\, x_s)^2, \nonumber\\
f_ 3 (x_s) & = & 4\big[-1 + 2\chi + \alpha (-1 - 23\chi + 14\chi^2) + \alpha^2 (37 + 36\chi - 
       147\,\chi^2 + 70\,\chi^3) + \alpha^3\, (-163 + 257\,\chi + 21\,\chi^2 - 301\,\chi^3 + 
        98\,\chi^4)\big],\nonumber\\
f_ 4 (x_s) & = & \big[-1 + 2\chi + \alpha (-1 - 23\chi + 14\chi^2) + \alpha^2 (37 + 36\chi - 
       147\chi^2 + 70\chi^3) + \alpha^3 (-163 + 257\chi + 21\chi^2 - 301\chi^3 + 
        98\chi^4)\big], \nonumber\\
f_ 5 (x_s) & = & 5 - 7\chi +  x_s + \sqrt {\Psi_ 3 (x_s)}\, \sqrt {\Psi_ 2 (x_s)} + \alpha\big[
 7 + 7\chi^2 - 5 x_s + 7\chi (2 + x_s)\big], \nonumber\\
 F_ 1 (x_s) & = & - 7 \sqrt{7} - 7 \sqrt{7} \, \chi^3 + (6 - 14 \, \alpha + 2 \, \alpha^3\, (7 + 13 \, \alpha)\, x^4_s  + 4 \, \alpha^4 \, x^5_s  )\, F (x_s) \sqrt{\Psi_ 2 (x_s)};, \nonumber \\
F_ 2 (x_s) & = &11 \sqrt{7} + (10 \, \alpha + 96 \, \alpha^2 - 42 \,\alpha^3 )\,F (x_s)\sqrt{\Psi_ 2 (x_s)},~~~F_ 3 (x_s)  = \sqrt{7}+ (18\, \alpha^2 + 84 \, \alpha^3 -  14 \, \alpha^4 )\, F (x_s) \,\sqrt {\Psi_ 2 (x_s)}, \nonumber \\
F_ 4(x_s) & = &3 \sqrt {7} + 2 (1 + 22 \, \alpha - 21\, \alpha^2) F (x_s)  \sqrt {\Psi_ 2 (x_s)},~~~F_ 5 (x_s) = 13 \sqrt {7} + (84 \, \alpha^2 + 84 \, \alpha^3 ) F (x_s) \, \sqrt {\Psi_ 2 (x_s)},\nonumber\\
F_ 6 (x_s) &=& 3 \sqrt {7} + 3 \sqrt {7} x_s + (2 \alpha + 6 \, \alpha^2\,  x_s  + 6 \, \alpha^3\,  x^2 _s  + 2 \, \alpha^4\,  x^3 _s ) F (x_s)  \sqrt {\Psi_ 2 (x_s)} \nonumber \\
F_ 7 (x_s) & = & 18 x_s  \sqrt {7} +21\, \sqrt {7} + \big[28\, \alpha^3 (3 + \alpha)  x^3_s + 28\, \alpha^4  x^4_s  + 28\, \alpha + 
2 x_s (14 \, \alpha + 42 \, \alpha^2 )\big] F (x_s) \sqrt {\Psi_ 2 (x_s)}. \nonumber
\end{eqnarray}

\section{The expressions for used coefficients in Eqs.(\ref{51})  and (\ref{EOSpt}) are as follows:}\label{B} 
Here $\tilde{\rho}$, $\tilde{\rho}_c$ and  $\tilde{\rho}_s$ are effective energy density, effective central density and effective surface density respectively. \\

\begin{eqnarray}
\rho_1&=&\bigg[\frac{32 - 84\, \tilde{\rho}_1\,  (1 + \chi)}{84\,\tilde{\rho}_1} +\frac{
 8\, 2^{2/3}\, \big[(8 + 63\, \tilde{\rho}_1 \, \chi)+ \big(\psi_ 1(\rho_ 1)\big)^{2/3}\big]}{84\,\tilde{\rho}_1\,\big(\psi_ 1 (\rho_ 1)\big)^{1/3}} \bigg],~~\tilde{\rho}_{1}=\frac{72\,\tilde{\rho}}{7\,(1+\chi)^2\,\tilde{\rho}_c}\nonumber\\
 \rho_{1s}&=&\bigg[\frac{32 - 84\, \tilde{\rho}_{1s}\,  (1 + \chi)}{84\,\tilde{\rho}_{1s}} +\frac{
 8\, 2^{2/3}\, \big[(8 + 63\, \tilde{\rho}_{1s} \, \chi)+ \big(\psi_ 1(\rho_ {1s})\big)^{2/3}\big]}{84\,\tilde{\rho}_{1s}\,\big(\psi_ 1 (\rho_ {1s})\big)^{1/3}} \bigg],~~\tilde{\rho}_{1s}=\frac{72\,\tilde{\rho_s}}{7\,(1+\chi)^2\,\tilde{\rho}_c}\nonumber\\
\psi_ 1 (\rho_ {1}) &=& 32+378\, \tilde {\rho}_{1}\,\chi+(2646 +  1323 \,\chi - 1323\,\chi^2)\,\tilde{\rho}^2_{1} + \sqrt{ \big[32 + 378\,\tilde{\rho}_ {1} \, \chi +1323\, \tilde {\rho}^2_ {1} \, (2 + \chi - \chi^2)\big]^2-2\, (8 + 63 \, \tilde {\rho}_ {1}\,\chi)^3}\nonumber\\
\psi_ 1 (\rho_ {1s}) &=& 32+378\, \tilde {\rho}_{1s}\,\chi+(2646 +  1323 \,\chi - 1323\,\chi^2)\,\tilde{\rho}^2_{1s} + \sqrt{ \big[32 + 378\,\tilde{\rho}_ {1s} \, \chi +1323\, \tilde {\rho}^2_ {1s} \, (2 + \chi - \chi^2)\big]^2-2\, (8 + 63 \, \tilde {\rho}_ {1s}\,\chi)^3}\nonumber\\
G(\rho_{1s})&=&\frac{2 (1 + \alpha\, \rho_{1s})^3\,\sqrt{\Psi_2(\rho_{1s})}\, \big[3 + \rho_{1s}-\alpha\,\Psi_7(\rho_{1s})\,\big]}{\big[F_ 1 (\rho_{1s}) + \rho^2_{1s}\, F_ 2 (\rho_{1s}) + \rho^3_{1s} \, F_ 3 (\rho_{1s}) + F_{11}(\rho_{1s})\big]},~~~H(\rho_{1s})=\frac{1}{\Psi_1(\rho_{1s})\,\big[1+G(\rho_{1s})\,\,F(\rho_{1s})\big]}\sqrt{\frac{7-10\,\rho_{1s}-\rho^2_{1s}}{7(1+\chi+\rho_{1s})^2}} ,\nonumber\\ 
\Psi(\rho_1)&=&\Psi_{1}(\rho_1)\Bigg[H(\rho_{1s})+L(\rho_{1s})\, F(\rho_1)\Bigg],~~ F(\rho_1)=\frac{\sqrt{7}\,\sqrt{\Psi_ 2 (\rho_1)} \big[f_ 1 (\rho_1) + f_ 2 (\rho_1)\big]}{3\, \Psi_ 3 (\rho_1)\,\Psi_1(\rho_1)\,\sqrt{\Psi_1(\rho_1)}} +\frac{\sqrt{7}\,f_3 (\rho_1) }{(\Psi _ 3 (\rho_1))^{7/2}}\,\ln\bigg[\frac {-a (\Psi_3 (\rho_1))^{5/2}\, f_ 5 (\rho_1)} { 2 \, f_ 4 (\rho_1)\, (1 +\alpha\,\,\rho_1)}\bigg],\nonumber\\
F(\rho_{1s})&=& \frac{\sqrt{7}\,\sqrt{\Psi_ 2 (\rho_{1s})} \big[f_ 1 (\rho_{1s}) + f_ 2 (\rho_{1s})\big]}{3\, \Psi_ 3 (\rho_{1s})\,\Psi_1(\rho_{1s})\,\sqrt{\Psi_1(\rho_{1s})}} +\frac{\sqrt{7}\,f_3 (\rho_{1s}) }{(\Psi _ 3 (\rho_{1s}))^{7/2}}\,\ln\bigg[\frac {-a (\Psi_3 (\rho_{1s}))^{5/2}\, f_ 5 (\rho_{1s})} { 2 \,f_ 4 (\rho_{1s})\, (1 + \alpha\,\rho_{1s})}\bigg],\nonumber\\
L(\rho_{1s})&=&G(\rho_{1s})\,H(\rho_{1s}),~~~
F_{11}(\rho_{1s})=\rho_{1s}\, F_ 4 (\rho_{1s}) - 7 \, \chi^2 \, F_ 6 (\rho_{1s}) - \chi \, \big[F_ 7 (\rho_{1s}) + \rho^2_{1s} \, F_ 5 (\rho_{1s})\big],\nonumber\\
\Psi_ 1 (\rho_1) &=&(1 + \alpha\, \rho_1)^2,~~~~\Psi_ 2 (\rho_1) = \big[7 + 7 \chi^2 - 10 \rho_1 - \rho_1^2 + 14 \chi (1 + \rho_1)\big],~~\Psi_ 3 (\rho_1) = \big[ 7 \alpha^2 (1 + \chi)^2 - 2 \alpha (-5 + 7 \chi)-1\big],\nonumber\\ 
\Psi_ 1 (\rho_{1s}) &=& (1 + \alpha\, \rho_{1s})^2, ~~ \Psi_ 2 (\rho_{1s}) = \big[ 7 + 7\chi^2 - 10 \rho_{1s} - \rho_{1s}^2 + 14\chi (1 + \rho_{1s})\big], ~~ \Psi_ 3 (\rho_{1s}) = \big[7\alpha^2 (1 + \chi)^2 - 2\alpha (-5 + 7\chi)-1\big], \nonumber\\
\Psi_ 4 (\rho_ 1) & = &\frac {\sqrt {7}\, L(\rho_{1s})\, (1 + \chi + \rho_ 1)} {\Psi_ 1 (\rho_ 1)\, \Psi (\rho_1)\, \sqrt {\Psi_ 2 (\rho_ 1)}} + \frac {2\, \alpha} {\sqrt {\Psi_ 1(\rho_ 1)}}, ~~~\Psi_ 5 (\rho_ 1) = \alpha^2\, [5 - \chi + \rho_ 1 + 2\alpha\, (-6 + 7\, \rho_ 1 + \rho_ 1^2 - \chi\, (6 + 5\, \rho_ 1))], \nonumber \\ 
\Psi_ 6 (\rho_ 1) & = &\alpha^2\, [\, 7 + 7\chi^3 - 15\, \rho_ 1 - 2\, \rho_ 1^2 +  21\, \chi^2\, (1 + \rho_ 1)\,],~~\Psi_7(\rho_{1s})=\big[7+7 \,\chi^2-13 \, \rho_{1s}-2\,\rho^2_{1s}+14\,\chi\,(1 + \rho_{1s})\big],~~~\nonumber\\ 
f_ 1 (\rho_1) &=& (1 - \alpha - \alpha\,\chi) \Psi^2_ 3 (\rho_1) + \Psi_ 3 (\rho_1)\, \big[-1 + 6 \alpha \chi + \alpha^2 (-23 - 16 \chi + 7 \chi^2)\big] (1 + \alpha \rho_1),\nonumber\\
f_ 2 (\rho_1) &=& \big[1 + \alpha (7 - 23 \chi) + \alpha^2 (63 + 174 \chi - 105 \chi^2) + \alpha^3 (-359 + 201 \chi + 399 \chi^2 - 161 \chi^3)\big]\,(1 + \alpha\,\rho_1)^2,\nonumber\\
f_ 3 (\rho_1) &=& 4 \big[-1 + 2 \chi + \alpha (-1 - 23 \chi + 14 \chi^2) + \alpha^2 (37 + 36 \chi - 147 \chi^2 + 70 \chi^3) + \alpha^3 (-163 + 257 \chi + 21 \chi^2 - 301 \chi^3 + 98 \chi^4)\big],\nonumber\\
f_ 4 (\rho_1) &=& \big[-1 + 2 \chi + \alpha (-1 - 23 \chi + 14 \chi^2) + \alpha^2 (37 + 36 \chi - 147 \chi^2 + 70 \chi^3) +
    \alpha^3 (-163 + 257 \chi + 21 \chi^2 - 301 \chi^3 + 98 \chi^4)\big],\nonumber
    \end{eqnarray}  
\begin{eqnarray}
f_ 5 (\rho_1) &=& 5 - 7 \chi + \rho_1 + \sqrt{\Psi_ 3 (\rho_1)}\, \sqrt{\Psi_ 2 (\rho_1)} + \alpha \big[7 + 7 \chi^2 - 5 \rho_1 + 7 \chi (2 + \rho_1)\big],\nonumber\\
f_ 1 (\rho_{1s}) & = & (1 - \alpha - \alpha\, \chi)\Psi^2 _ 3 (\rho_{1s}) + \Psi_ 3(\rho_{1s})\, \big[-1 +
    6\,\alpha\,\chi + \alpha^2\, (-23 - 16\chi + 7\chi^2)\big] (1 + \alpha \rho_{1s}), \nonumber\\
f_ 2 (\rho_{1s})& = &  \big[1 + \alpha (7 - 23\chi) + \alpha^2 (63 + 174\chi - 105\chi^2) + \alpha^3 (-359 + 201\chi + 399\chi^2 - 161\chi^3)\big]\, (1 + \alpha\, \rho_{1s})^2, \nonumber\\
f_ 3 (\rho_{1s}) & = & 4\big[-1 + 2\chi + \alpha (-1 - 23\chi + 14\chi^2) + \alpha^2 (37 + 36\chi -
       147\,\chi^2 + 70\,\chi^3) + \alpha^3\, (-163 + 257\,\chi + 21\,\chi^2 - 301\,\chi^3 +
        98\,\chi^4)\big],\nonumber\\
f_ 4 (\rho_{1s}) & = & \big[-1 + 2\chi + \alpha (-1 - 23\chi + 14\chi^2) + \alpha^2 (37 + 36\chi -
       147\chi^2 + 70\chi^3) + \alpha^3 (-163 + 257\chi + 21\chi^2 - 301\chi^3 +
        98\chi^4)\big], \nonumber\\
f_ 5 (\rho_{1s}) & = & 5 - 7\chi +  \rho_{1s} + \sqrt {\Psi_ 3 (\rho_{1s})}\, \sqrt {\Psi_ 2 (\rho_{1s})} + \alpha\big[
 7 + 7\chi^2 - 5 \rho_{1s} + 7\chi (2 + \rho_{1s})\big], \nonumber\\
 F_ 1 (\rho_{1s}) & = & - 7 \sqrt{7} - 7 \sqrt{7} \, \chi^3 + (6 - 14 \, \alpha + 2 \, \alpha^3\, (7 + 13 \, \alpha)\, \rho^4_{1s}  + 4 \, \alpha^4 \, \rho^5_{1s}  )\, F (\rho_{1s}) \sqrt{\Psi_ 2 (\rho_{1s})};, \nonumber \\
F_ 2 (\rho_{1s}) & = &11 \sqrt{7} + (10 \, \alpha + 96 \, \alpha^2 - 42 \,\alpha^3 )\,F (\rho_{1s})\sqrt{\Psi_ 2 (\rho_{1s})},~~~F_ 3 (\rho_{1s})  = \sqrt{7}+ (18\, \alpha^2 + 84 \, \alpha^3 -  14 \, \alpha^4 )\, F (\rho_{1s}) \,\sqrt {\Psi_ 2 (\rho_{1s})}, \nonumber \\
F_ 4(\rho_{1s}) & = &3 \sqrt {7} + 2 (1 + 22 \, \alpha - 21\, \alpha^2) F (\rho_{1s})  \sqrt {\Psi_ 2 (\rho_{1s})},~~~F_ 5 (\rho_{1s}) = 13 \sqrt {7} + (84 \, \alpha^2 + 84 \, \alpha^3 ) F (\rho_{1s}) \, \sqrt {\Psi_ 2 (\rho_{1s})},\nonumber\\
F_ 6 (\rho_{1s}) &=& 3 \sqrt {7} + 3 \sqrt {7} \rho_{1s} + (2 \alpha + 6 \, \alpha^2\,  \rho_{1s}  + 6 \, \alpha^3\,  \rho^2_{1s}  + 2 \, \alpha^4\,  \rho^3_{1s} ) F (\rho_{1s})  \sqrt {\Psi_ 2 (\rho_{1s})} \nonumber \\
F_ 7 (\rho_{1s}) & = & 18 \rho_{1s}  \sqrt {7} +21\, \sqrt {7} + \big[28\, \alpha^3 (3 + \alpha) \rho^3_{1s} + 28\, \alpha^4  \rho^4_{1s}  + 28\, \alpha +
2 \rho_{1s} (14 \, \alpha + 42 \, \alpha^2 )\big] F (\rho_{1s}) \sqrt {\Psi_ 2 (\rho_{1s})}. \nonumber 
\end{eqnarray}  

\end{widetext}


\end{document}